\documentclass[apj]{emulateapj}
\usepackage[alpha,z]{tmsfnts}
\def\gtsim {\lower .1ex\hbox{\rlap{\raise .6ex\hbox{\hskip .3ex
        {\ifmmode{\scriptscriptstyle >}\else
                {$\scriptscriptstyle >$}\fi}}}
        \kern -.4ex{\ifmmode{\scriptscriptstyle \sim}\else
                {$\scriptscriptstyle\sim$}\fi}}}
\newcommand{\etal}{{~et al.~}}

\newcommand{\beq}{\begin{equation}}
\newcommand{\eeq}{\end{equation}}

\newcommand{\lap}{\lesssim}

\def\kpc{\ {\rm kpc}}

\slugcomment{Accepted for Publication in ApJ {March 20, 2007, v658n 1}}

\shorttitle{Relaxation processes in mergers}
\shortauthors{Valluri \etal }

\begin{document}

\title{On relaxation processes in collisionless mergers}

\author{Monica Valluri,\altaffilmark{1,2}
Ileana M. Vass, \altaffilmark{3,2}
Stelios Kazantzidis,\altaffilmark{2,6} 
Andrey V. Kravtsov\altaffilmark{1,2,4} and
Courtlandt L.Bohn \altaffilmark{5}
}
\altaffiltext{1}{Department of Astronomy and Astrophysics, University of Chicago, 5640 S. Ellis
Avenue, Chicago, IL 60637 \\
{\tt valluri@kicp.uchicago.edu}
}
\altaffiltext{2}{Kavli Institute for Cosmological Physics, The University of Chicago, Chicago, IL
60637}
\altaffiltext{3}{Department of Astronomy, University of Florida, Gainesville, FL 32611
}
\altaffiltext{4}{Enrico Fermi Institute, The University of Chicago, Chicago, IL
60637}
\altaffiltext{5}{Department of Physics, Northern Illinois University, DeKalb, IL 60115}
\altaffiltext{6}{Kavli Institute for Particle Astrophysics and Cosmology, Department of
Physics,  Stanford University,  Stanford, CA 94309}

\begin{abstract}
We analyze $N$-body simulations of halo mergers to investigate the
mechanisms responsible for driving mixing in phase-space and the
evolution to dynamical equilibrium. We focus on mixing in energy and
angular momentum and show that mixing occurs in step-like fashion
following pericenter passages of the halos.  This makes mixing during
a merger unlike other well known mixing processes such as phase mixing
and chaotic mixing whose rates scale with local dynamical time.  We
conclude that the mixing process that drives the system to equilibrium
is primarily a response to energy and angular momentum redistribution
that occurs due to impulsive tidal shocking and dynamical friction
rather than a result of chaotic mixing in a changing potential. We
also analyze the merger remnants to determine the degree of mixing at
various radii by monitoring changes in radius, energy and angular
momentum of particles. We confirm previous findings that show that the
majority of particles retain strong memory of their original kinetic
energies and angular momenta but do experience changes in their
potential energies owing to the tidal shocks they experience during
pericenter passages. Finally, we show that a significant fraction of
mass ($\approx 40\%$) in the merger remnant lies outside its formal
virial radius and that this matter is ejected roughly uniformly from
all radii outside the inner regions. This highlights the fact that
mass, in its standard virial definition, is not additive in
mergers. We discuss the implications of these results for our
understanding of relaxation in collisionless dynamical systems.

\end {abstract}


\section{Introduction}

Almost four decades have passed since ``violent relaxation'' was first
proposed by \citet{lyndenbell67} as a mechanism to explain the
remarkable regularity and smoothness of the light distributions in
elliptical galaxies. Several authors had noted that the time scale for
two-body relaxation \citep{chandrasekhar43a} was several orders of
magnitude too long to be relevant for the evolution of galaxies
\citep{zwicky39}. Lynden-Bell derived the distribution function that
would result from rapid fluctuations in the potential of the galaxy,
such as those that might occur during gravitational collapse or
following a merger. The process of violent relaxation outlined by
Lynden-Bell produces a smooth distribution function as a result of the
gravitational scattering of particles by the time
dependent global gravitational potential. He argued that this
process could result in a mass-independent and smooth final
distribution of particles in phase-space.

Another process that was considered by \citet{lyndenbell67} is ``phase
mixing'' - a process by which an initially compact ensemble of
non-interacting phase-space points in a time-independent potential can
be stretched into a long ribbon in phase-space as a result of small
differences in the initial energies (or angular momenta) of particles
in the ensemble.  Phase mixing conserves the fine grained distribution
function but in a coarse grained sense it produces smooth/relaxed
looking distributions on timescales that depend only on the range of
values of the orbital integrals (or actions) in the ensemble
\citep{BT}.

In the last decade, a new type of mixing process, which occurs due to
the presence of large fractions of chaotic orbits in a galactic
potential (such as when box orbits are scattered by a central
supermassive black hole in a triaxial galaxy), has been investigated
\citep{kandrup_mahon94,MV96,habib97,KPS,kandrupSiopis03}. These studies
have shown that microscopic ensembles of iso-energetic test particles
which occupy an infinitesimal volume of phase-space surrounding a
chaotic orbit can evolve rapidly to a near-invariant distribution that
uniformly covers the energy surface available to the ensemble.  This
process of diffusion to a near-invariant distribution is termed
``chaotic mixing'' and has been shown to occur in a variety of
galaxy-type potentials. These authors argued that in a triaxial galaxy
with a supermassive central black hole, a significant fraction of
orbits are chaotic, and consequently, this mixing process could drive
secular evolution to more axially symmetric distributions.

The physically interesting aspect of this mixing process is that the
evolution to a near-invariant distribution has been shown to occur on
timescales of order 5 to 100 orbital crossing times in strongly
chaotic systems making it a potentially important process for the
evolution of galaxies.  This was demonstrated by
\citet{merritt_quinlan98} who employed $N$-body simulations of
triaxial elliptical galaxies containing growing supermassive central
black holes and found the triaxial figures to evolve to more
axisymmetric shapes.

Like phase mixing, chaotic mixing conserves the fine-grained
phase-space density.  Unlike phase mixing, its rate does not depend on
the initial spread in the energy distribution of particles in the
ensemble, but on how strongly chaotic the orbits in the ensemble are
\citep{MV96}.

However two concerns remain regarding these
earlier experiments. First the rate of evolution to the invariant
measure depends on the properties of the ensemble of particles. For
ensembles near a ``sticky'' or weakly chaotic orbit, a near-invariant
distribution may not be attained in the lifetime of the galaxy
\citep{MV96}. Second, the majority of the numerical studies which
found strong chaotic mixing and short time scales for evolution were
not carried out in self-consistent potentials leading to concerns that
in self-consistent systems such mixing could be self-limiting.

Since all orbits in a time-dependent potential with long-lived
oscillations can, in principle, exchange energy with the potential, a
large fraction of the orbits will not conserve the energy integral. In
such systems previous authors have found that a significant fraction
of orbits can be chaotic \citep{KVS}. This raises the possibility that
the underlying physical process driving relaxation in violently
fluctuating potentials could be a consequence of chaotic mixing. The
first step in making this connection was studied by \citet{KVS} and
\citet{TK}, who demonstrated that large fractions of chaotic orbits
arise in potentials subjected to long-duration (damped) periodic
oscillations. These authors found that the fraction of chaotic orbits
depended on the frequency as well as the amplitude of the oscillations
in the potential and attributed the chaos to a ``broad
parametric resonance''.

In this paper we analyze  the mechanisms driving the approach to 
equilibrium following the collisionless
merger of two self-gravitating systems.  Our principal interest is in
understanding whether changes in the fine-grained distribution
function (e.g., changes in the angular momenta and energies of nearby
particles) occur as a result of exponential instabilities of orbits or
some more straightforward energy exchange process which transfers
kinetic energy of the merging halos to the internal energies of
particles.

We  consider two dynamical heating effects that are not
usually discussed in this context.  The first is the effect of
compressive tides that arise when one collisionless gravitating system
passes through another on a time scale that is short compared to the
internal dynamical time of the infalling system. Such tides can
impulsively heat particles as a result of the transient deepening
of the net potential. Impulsive compressive tidal shocks have been
studied previously and are well known to produce large changes in the
internal structure and evolution of globular clusters
\citep{spitzerChevalier73,gnedin99}, galaxies orbiting within a
cluster potential \citep{valluri93}, and DM subhalos orbiting within
larger hosts \citep{kravtsov_gnedin_etal04,kazantzidis_etal04b}.

The second process is dynamical friction \citep{chandrasekhar43a},
which is the primary mechanism that causes the formation of a single
remnant when two collisionless halos merge. When collisionless halos
merge, dynamical friction transfers their relative linear and angular
orbital momenta to the individual particles. The original version of
the \citet{chandrasekhar43a} dynamical friction theory applied to the deceleration of a
massive star  moving in an infinite homogenous isotropic stellar distribution.
 However several studies have shown that with minor
modifications the theory is applicable in systems that resemble real
galaxies and clusters \citep{White76,binney77,velazques_white99}.
Dynamical friction is found to be enhanced in systems with
inhomogenous density profiles (such as cuspy profiles \citeauthor{Delpopolo03} \citeyear{Delpopolo03}), as compared with
systems with uniform homogenous distributions.  Recent $N$-body
studies of live satellites in extended $N$-body galaxies \citep{
jiang_binney00, fujii_etal06} have shown that the timescale for
dynamical friction is shorter than that predicted by Chandrasekhar due
to two additional components of drag force: the tidal torque from the
leading tidal tail and an enhanced wake due to particles that are no
longer bound to the satellite but still trail behind it in the same orbit.

It is important to emphasize at the outset, that the terms ``mixing''
and ``relaxation'' are often used rather loosely (and often
interchangeably) in the astrophysics literature to mean the approach of
a gravitating system to global as well as small-scale equilibrium. The
most restrictive definition of ``relaxation'' is a process which leads
to the complete ``loss of memory of initial conditions'' or loss of
correlations between initially nearby particles in phase space. The
large changes in integrals of motion required for such relaxation only
occur as a result of two-body interactions \citep{chandrasekhar43a}
and as a result of ``violent relaxation''. Both processes cause
changes in the fine-grained distribution function. In
contrast phase mixing and chaotic mixing in time-independent
potentials, under the right conditions, can lead to large changes in
correlations but conserve integrals of motions (if they exist) and
conserve the fine-grained distribution function. In this paper we
focus our attention on loss of correlations in phase space
accompanying a collisionless merger and use the term ``mixing'' to
connote the approach to macroscopic equilibrium and the accompanying
change in the fine-grained distribution function (even though this is
quite small).

According to the prevailing cosmological model for structure
formation, galaxies and dark matter (DM) halos in the Universe form
hierarchically via multiple mergers. Numerous minor mergers and
several major mergers of subhalos eventually form the dark halos at
$z=0$. Thus, the merger history of cosmological structures is complex,
and DM halos are predicted to abound in substructure and have mass
profiles that may still be evolving today.  A detailed analysis of the
physics of gravitational relaxation, with the specific goal of
understanding what processes drive the DM and stellar component to
equilibrium, can be more readily done using dissipationless controlled
merger simulations.

Several aspects of the universality in the structure and properties of
dark matter halos have lead to new interest in the detailed physical
processes driving gravitational relaxation. The discovery of
``universal'' dark matter halo profiles resulting from cosmological
$N$-body simulations
\citep[e.g.,][]{dubinski_carlberg91,NFW,navarro_frenk_white97}, the
more recent realization that these profiles have only a ``nearly universal''
form with a correlation between the density profile's shape parameter
and the halo mass \citep{merritt_etal05, merritt_etal06} as well as the 
finding that phase-space density profiles of these halos have
power-law distributions over about 2.5 orders of magnitude in radial
extent \citep{taylorN01,Arad04,ascasibar_binney05} all point to a universality in
the processes driving structure formation. In particular,
violent relaxation and the statistical distribution functions that
result from this process are being actively investigated
\citep{aradLB05, aradJ05, dehnen05}. Several studies of the remnant
density profiles have shown that the remnants retain a surprisingly
strong memory of their initial density profiles despite the strength
of the relaxation process \citep{barnes96,boylan-kolchin_ma04,
kazantzidis_etal06}.

Several authors have criticized Lynden-Bell's \citeyear{lyndenbell67}
original suggestion that it is potential fluctuations that occur
during gravitational collapse that drive the system to equilibrium, on
the grounds that time dependence of the potential alone merely causes
particles to be relabeled in energy and causes no mixing of nearby
particles in energy and angular momentum.  It has been therefore
argued that a process like chaotic mixing is actually responsible (at
the microscopic level) for driving the evolution of the system to
equilibrium \citep{MV96,KVS}. 

An analysis of the merging process can
help us better understand the physical mechanisms that drive the evolution to
equilibrium and cause the redistribution of nearby particles in
phase-space.
In this investigation we restrict ourselves to the study of mixing in positions, velocities,  energies and angular momenta of particles. The goals of this study
are: (a) to determine whether the relaxation of a remnant to an equilibrium
distribution following a collisionless  $N$-body merger is driven by chaotic 
mixing that is believed to
take place in time-independent potentials (b) to quantify the
degree of relaxation/mixing in various phase-space parameters that
occurs during the merger by examining the remnants of major mergers of
DM halos, and (c) to obtain a better understanding of why steep cusps are
as robust as they are.  Both the time-dependent nature of the problem,
as well as the fact that $N$-body systems exhibit exponential
sensitivity to initial conditions (as discussed in greater detail in
\S~\ref{sec:micro}), make the first of the above goals particularly
challenging.

The paper is organized as follows. In \S~2 we describe the $N$-body
experiments that were conducted and define several new measures that
we use to quantify the rate of separation of nearby orbits and the
degree of ``mixing'' or ``relaxation'' in phase-space.  In \S~3 we
discuss the results of the application of our measures of mixing to
the merger experiments by considering the separation in various
phase-space variables of pairs of nearest neighbors in phase-space
(\S~\ref{sec:separation}) as well as by monitoring the rate of mixing
of ensembles of nearby particles (\S~\ref{sec:mixing}).  In \S~4 we
examine how particles are redistributed in phase-space following the
merger. In \S~5 we summarize our results and discuss the implications
of our findings for understanding relaxation during mergers and the
mixing that it produces.

\section{Numerical Methods}

We investigate the mechanisms responsible for violent relaxation by
analyzing $N$-body simulations of binary mergers between dark matter
halos. The halo models follow density profiles that are described by
the general $(\alpha,\beta,\gamma)$ spherical density law (eq 1.)
\citep[e.g.,][]{Hernquist90,zhao96}, where $\gamma$ denotes the asymptotic inner
slope of the profile, $\beta$ corresponds to the outer slope, and 
$\alpha$ determines the sharpness of the transition between the inner and outer
profile.  
\begin{equation} 
\rho(r) = 
{\frac{\rho_s}{(r/r_s)^{\gamma}[1+(r/r_s)^{\alpha}]^{(\beta-\gamma)/\alpha}}}
 \qquad (r\leq r_{\rm vir})  
\end{equation}

We consider two main halo models specified by particular choices of
the parameters $\alpha$, $\beta$, and $\gamma$. The first model
follows the \citet[][hereafter NFW]{NFW} profile (with $(\alpha, \beta,
\gamma)=(1,3,1)$), while the second model 
(with $(\alpha, \beta, \gamma)=(2,3,0.2)$) corresponds to a profile with
a shallower inner
slope. $N$-body halo models are constructed using the exact
phase-space distribution function under the assumptions of spherical
symmetry and an isotropic velocity dispersion tensor
\citep{kazantzidis_etal04a}. Each of the initial DM halos has a virial
mass of $M_{\rm vir}=10^{12}M_{\odot}$ implying a virial radius of
$r_{\rm vir} \simeq 256.7 \kpc$, and a concentration of $c=12$,
resulting in a scale radius of $r_{\rm s} \simeq 21.4 \kpc$. It is
worth emphasizing that the adopted value of $M_{\rm vir}$ serves
merely practical purposes and does not imply anything special about
the particular choice of mass scale.  Hence, our conclusions can be
readily extended to mergers between equal-mass systems of any mass
scale.

We analyze three different merger experiments which were conducted
with the multi-stepping, parallel, tree $N$-body code PKDGRAV
\citep{stadel01}. PKDGRAV uses a spline softening length, such that
the force is completely Keplerian at twice the quoted softening
length, and multi-stepping based on the local acceleration of
particles. The first experiment followed the encounter of two NFW
halos (referred to hereafter as run Bp1), 
while in the second experiment an NFW halo merged with a halo
having an inner slope of $\gamma = 0.2$ (referred to hereafter as run hBp1). 
 The initial halo models  in these two runs were
sampled with $N = 2 \times 10^5$ particles and forces were softened
with a spline gravitational softening length equal to $\epsilon =
1.5\kpc$. In order to minimize any concern that our results might be
compromised by numerical resolution effects, we analyzed an additional
merger experiment between two NFW halos increasing the mass resolution
by a factor of $10$ and scaling down the softening lengths according
to $\epsilon \propto N^{-1/3}$ (hereafter referred to as run HRBp).  Initial conditions for binary mergers
were generated by building pairs of halo models and placing them at a
distance equal to twice their virial radii.  In this study we only
discuss mergers of systems on parabolic orbits owing to the fact that
this particular orbital configuration is the most typical of merging
halos in cosmological simulations
\citep[e.g.,][]{zentner_etal05}. These merger simulations (labeled Bp1,hBp1, and HRBp) were
analyzed and described in greater detail in \citet{kazantzidis_etal06}. Extensive convergence 
tests carried out by  \citet{kazantzidis_etal06} indicate that isolated halo models did not deviate from their original distribution functions on timescales as long as ~100 crossing times even  at small radii ($r \sim \epsilon$).

In order to distinguish the mixing that results from the exponential
instability of the $N$-body problem (see \S~\ref{sec:micro}) from the
effects of the mixing resulting during the merger, we also performed
$N$-body simulations of the evolution of a spherical isolated NFW
halo. All orbits in an isolated spherical halo are expected (in the
smooth potential or large $N$ limit) to be ``regular orbits'' that
conserve four isolating integrals of motion. This simulation therefore
serves as an important control experiment with which to compare the evolution in
the merger simulations.
%
\begin{figure}
\epsscale{0.75}
\plotone{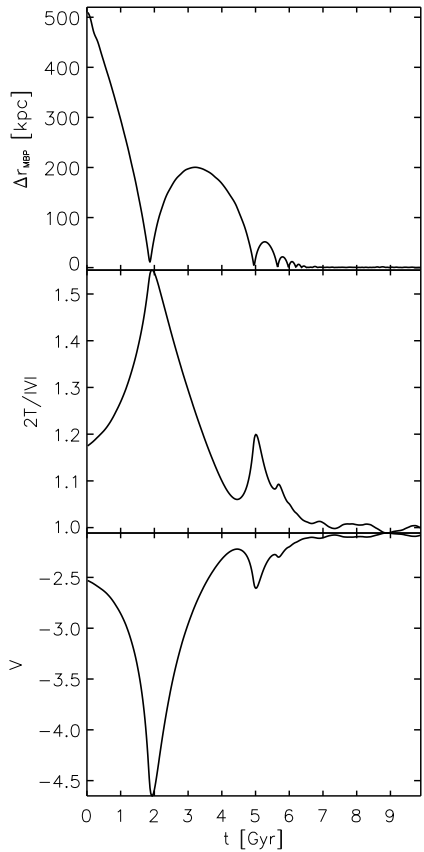}
\caption{Top: The separation between the most-bound particles
of the two merging NFW halos (run Bp1) as a function of
time; Middle: time evolution of the virial ratio
$2T/|V|$; Bottom: time evolution of the total potential energy of the system $V$.}
\label{fig:mbpvir}
\end{figure}

\subsection{Macroscopic Evolution of Merger Remnants}

A self-gravitating distribution of particles that is out of
equilibrium will experience potential fluctuations, and the potential
and kinetic energies of particles will oscillate back and forth. This
behavior is governed by the time-dependent virial theorem:
\begin{equation}
{\frac{1}{2}}\ddot{I} = 2T + V
\end{equation}
where $I$ is the moment of inertia, $T$ is the total kinetic energy,
and $V$ is the total potential energy of the system of particles, with
all quantities defined with respect to the center of mass of the
system. For a time-independent gravitating system, $\ddot{I} = 0 = 2T
+ V$. A robust quantitative measure of how far a gravitating system is
from equilibrium can be obtained from the virial ratio $2T/|V|$, which
is unity for a system in global dynamical equilibrium.

In Figure~\ref{fig:mbpvir}, the top panel shows the separation of the
most-bound-particle (MBP) of each halo in the merger, the middle
panel shows the evolution of the virial ratio $2T/|V|$ and the lower panel shows the change in total potential energy of the system ($V$), as a function
of time.  We note that pericenter passages (seen as minima in the MBP
separation) correspond to maxima in the virial ratio and minimum in potential energy.
For $t \lap 7$ Gyr the virial ratio $2T/|V| $  and potential $V$ undergo strong
fluctuations  but remains close to unity thereafter. We refer to this
time $t \sim 7 $ Gyr to be the time when the system is {\it globally
relaxed}.

\subsection{Microscopic Chaos or the Miller Instability  \label{sec:micro}}

In real galaxies, there are two components of the gravitational
accelerations on particles, a component arising from the background
potential which is smoothly varying with time, and a second component
arising from the discrete nature of stars (and possibly dark-matter
particles). The importance of the discrete component is defined by the
two-body relaxation timescale \citep{chandrasekhar43a, BT} and is
larger than the age of the Universe in galaxy-sized
systems. Therefore, it is generally assumed that equilibrium and
non-equilibrium evolution of the phase-space density distributions in
galaxy-size objects can be studied in the mean-field limit, where the
collisionless Boltzmann equation can be applied.

The Lyapunov exponent \citep{LL92} is one of the most popular ways of
measuring the chaoticity of an orbit in a smooth potential. It
measures the rate of divergence of two infinitesimally nearby orbits
in phase-space. If the rate of divergence of phase-space points is
linear, then the Lyapunov exponent $\lambda = 0$ and the orbit is
termed {\it regular}. In a smooth equilibrium potential with
three-spatial dimensions such orbits will conserve at least three
isolating integrals of motion.  If, on the other hand, the rate of
divergence of the phase-space points is exponential, the orbit is
termed {\it chaotic}. Thus, the Lyapunov exponent defines the time scale on which
infinitesimally nearby trajectories in phase-space undergo exponential
separations in their phase-space coordinates.

However, as was first noted by \citet{miller64}, the $N$-body problem
is chaotic in the sense that the trajectory of the $6N$-dimensional phase-space
coordinate of the system exhibits exponential sensitivity toward small
changes in initial conditions. This exponential instability (referred
to in the literature as the ``Miller instability''; e.g.,
\citeauthor{hemsendorfM02} \citeyear{hemsendorfM02} or ``microchaos''
\citeauthor{sideris_kandrup02} \citeyear{sideris_kandrup02}) has been
investigated extensively in several studies over the past three
decades (see \citeauthor{merritt05} \citeyear{merritt05} for a
review). These studies have shown that the largest $N$-body Lyapunov
exponent $\lambda$ does not converge toward zero as $N$ increases
\citep{kandrupS91, goodmanHH93}, even for $N$-body realizations of
integrable potentials \citep{kandrupS01}.

 In fact more recently \citet{hemsendorfM02} have shown that the
direct $N$-body problem (with $N < 10^5$) is inherently chaotic and
that the degree of chaos, as measured by the rate of divergence of
nearby trajectories (with the Lyapunov exponent $\lambda$), increases
with increasing $N$ with characteristic $e$-folding time equal to
$\sim 1/20$ of the system crossing time (in the absence of softening). It has also been shown that
the mean $t_e$ ($e$-folding time) increases  with increasing particle number  in the case of softened potentials \citep{elzant02} so that discreteness effects are reduced as the softening
becomes comparable to interparticle separations
\citep{kandrup_smith91}.  

The exponential separation of particles in energy occurs on the Miller
instability timescale ($t_e$) which is much smaller than the two body
relaxation time ($t_r$) \citep{kandrup_mahon_smith94,
hut_heggie_01}. The divergence of particles in energy saturates after
a few $t_e$ and then varies with time on the standard two-body
relaxation timescale indicating that the Miller instability is not an
important process in changing the fine-grained distribution functions
of gravitating systems.
 
Despite the fact that particles separate exponentially in phase space,
as $N$ increases orbits in these systems become more and more regular,
acquiring orbital characteristics closer and closer to those in smooth
potentials, and the macroscopic separations of orbits saturate at
smaller and smaller distances
\citep{VM00,kandrupS01,Sideris04}. Additionally, decades of $N$-body
integrations have demonstrated that, in many ways, the behavior of
large-$N$ systems matches expectations derived from the collisionless
Boltzmann equation \citep[e.g.,][]{aarsethL75}.

The fact that in an $N$-body system the Lyapunov exponent will
primarily measure the ``Miller instability'' and not macroscopic chaos
(arising from the background potential) means that it is has limited
usefulness for measuring chaotic behavior in $N$-body systems. Other
measures of chaos, such as those that measure changes in orbital
characteristics (for instance, fundamental oscillation frequencies of
orbits; e.g., \citeauthor{laskar90} \citeyear{laskar90},
\citeauthor{laskar_etal92} \citeyear{laskar_etal92},
\citeauthor{valluri_merritt98} \citeyear{valluri_merritt98}), rely on
the ability to identify such frequencies from oscillations that last
between 30-100 orbital periods. Recently, a new technique based on
pattern recognition of orbital characteristics has been developed and
is applicable in time-dependent systems whose oscillations last 10-30
crossing times \citep{sideris06}.  To our knowledge, none
of the standard measures of chaos are
applicable to $N$-body orbits in potentials with non-periodic
potential fluctuations lasting only a few ($< 10$) crossing times.
This makes it necessary to define new ways of quantifying chaos and mixing.

\subsection{Definitions of mixing in $N$-body systems \label{sec:definitions}}

In this paper we estimate the degree of mixing at various stages of
the evolution of the merging systems (a) by considering {\em pairs} of
nearby particles in the $N$-body simulation and tracing their
separation in four different phase-space quantities and (b) by carrying
out {\it mixing experiments} that monitor the rate at which {\em
ensembles} of 1000s of nearby particles evolve with time and reach a
``near invariant'' distribution in energy and angular momentum.

\subsubsection{Separation of pairs of nearby particles \label{sec:pairs}}

Several previous studies of this type have focused on the
configuration space separation $\Delta r$ of a pair of infinitesimally
nearby test particles in a frozen $N$-body potential. These studies
picked pairs of non-self-gravitating test particles that are
arbitrarily nearby in phase-space \citep{VM00,kandrupS03}. Studies
that used live $N$-body simulations
\citep[e.g.,][]{kandrup_mahon_smith94} have compared pairs of orbits in
two different realizations of simulations where the initial conditions
had been perturbed by an infinitesimal amount.  Here we work with
orbits in a given $N$-body simulation and are therefore limited (by
the resolution of the simulations) in our ability to choose pairs of
particles that are very close in phase-space.

We select a large number ($ 1000$) of nearest-neighbor pairs in one of
the two merging $N$-body halos and follow their separations through
the entire merger process.  We have checked that when two identical
halos merge, the results were independent of the halo chosen to
perform the analysis.  To pick particle pairs we adopt the following
scheme:
\begin{enumerate}
\item{All the particles in a given halo are sorted in their separation $r$
from  the MBP of that halo, and the particles are binned in ten
radial shells of equal width, extending from the MBP to the initial
virial radius of the halo ($r_{\rm vir} \simeq 256.7 \kpc$). In  the analysis that
follows we compare behavior of particles in the 1st, 4th and 7th shell from the 
center of the potential. The outer radius of the first shell is just outside the scale 
radius, the 4th shell lies at the half mass radius and 7th shell 
 lies at the 3/4 mass radius of the initial  NFW halo.}
\item{A particle $P^*$ with separation $r_{\rm P}$ from the MBP is picked at random in
a given shell, and the 5000 nearest particles in configuration space
to the particle $P^*$ are selected.}
\item{The velocity separations $|v|$ of each of the 5000 particles
relative to particle $P^*$ are computed, and the 2500 particles with
the smallest $|v|$ separation from $P^*$ are selected.}
\item{The magnitudes of the separation of total angular momentum  $J$ of
each of the 2500 particles above relative to particle $P^*$ are
computed, and the 1250 particles with the smallest $J$ separation
from $P^*$ are selected.}
\item{The angle between the angular momentum ${\mathbf J}$ and the angular
momentum ${\mathbf J}_{P^*}$ is computed for each of the 1250 particles,
and the 625 particles with the smallest angle are selected.}
\item{From the remaining 625 particles, the particle with the total
energy $E = T + V$ closest to the particle $P^*$ is then picked as its
nearest neighbor.}
\end{enumerate}

In general, after an initial phase of exponential growth, the
separation of two nearby particles saturates and then oscillates about
some median value, due to small but finite differences in the initial
oscillation frequencies in the global potential \citep[e.g.,][]
{kandrupS03}.  To determine the behavior of a ``typical pair'' of
particles in a given shell, we first pick 1000 pairs of particles in
each shell. We then report the median value of the separation of that
phase-space quantity as a function of time for the 1000 particle
pairs. We find (as is well known) that the median is a more reliable
estimator than the mean separation of particles because each particle
has a small but finite probability of being ejected from the system as
a result of the merger, and such ejections cause a small fraction of
particles to form a long tail in the histogram of separations at each
time step.  Thus, for all quantities defined below it should be
implicitly understood that we report the median value of the parameter
for 1000 pairs of particles. Specifically, we monitor the following
quantities.
\begin{enumerate} 
\item{``$\ln(\Delta r)$'' -- the natural logarithm of the separation in the spatial coordinate $r$ of the median pair of particles.}
\item{``$\ln(|\Delta v|)$'' -- the natural logarithm of the absolute value of the  relative three dimensional velocity $v$ of the median pair of particles.}
\item{``$\Delta E$'' -- the separation in total energy of the median pair of particles.}
\item{``$\Delta J$'' -- the separation in the total amplitude of the angular momentum of the median pair of particles. (We pick pairs whose vector directions have angular separations of less than $90\deg$. As the particles separate, we track the separation in the amplitude of ${\mathbf J}$ but not the direction.)}
\end{enumerate}

The quantities defined above differ in one important aspect from
previous studies in which similar quantities have been used: all
particles in the pairs are self-gravitating (and not test particles)
and are picked from an initial distribution function that is
self-consistent with the spherical NFW potential. This implies that the typical
initial pair separation $\ln(\Delta r)$ of particles depends on their
location in the potential such that  pairs in the central
cusp will be much closer than the typical initial pair separation in
the outer regions.  We therefore often scale the quantities relative
to their values at $t=0$ Gyr. Moreover, especially when calculating
the velocity separation $\ln(\Delta v)$, we observe the mutual
gravitational attraction of nearby particles.
 
The quantities $\Delta E$ and $\Delta J$ provide additional insights
because in the case of an isolated $N$-body halo, both quantities are
expected to remain constant (to within numerical errors) for the
entire duration of the simulation (following the initial increase in
these quantities due to the Miller instability).  Since $E$ and
$\mathbf J$ are integrals of motion in equilibrium potentials, any
changes in the separation of these quantities in the case of the
merger reflects the influence of the time-dependent potential.  
Evolution with time of these
quantities for various simulations are described in
\S~\ref{sec:separation}.

\subsubsection{Mixing of ensembles in phase space \label{sec:ensembles}}

Another way to quantify the mixing rate is by monitoring the secular evolution of ensembles
of infinitesimally nearby test particles due to the presence of chaotic orbits. 
We conducted a series of mixing experiments where we
monitored the spread of an ensemble of 1000 nearby particles 
as a function of time in two phase-space
coordinates.  

Such experiments have been carried out previously to
study the mixing of microscopic ensembles of chaotic orbits in static,
non-linear potentials \citep{MABK95, MV96, KPS,
kandrupSiopis03}. These authors picked ensembles of orbits with
initial conditions that uniformly sampled an infinitesimally small
region of phase-space and measured the rate at which the ensemble
spreads and fills a region of configuration space.  Ensembles
associated with  strongly chaotic orbits were found to evolve to a
near-invariant distribution within 5-10 orbital crossing times, while
ensembles associated with weakly chaotic or ``sticky'' orbits (orbits
that lie close to resonance in phase-space) evolved on much longer
time scales and often did not reach an invariant
distribution. Ensembles associated with regular orbits only spread
slightly, due to phase mixing that resulted from the small differences
in their initial integrals of motion.

Once again the resolution of the simulations  determine how closely we can sample the phase-space.  We pick ensembles of the nearest 1000 particles about a given particle ($P^*$) in the higher resolution merger simulation only (run HRBp), following a
scheme similar to that for picking pairs of nearby particles ( \S~\ref{sec:pairs} ) to obtain
the 1000 nearest particles to $P^*$.

Configuration-space coordinates have traditionally been used to study
chaotic mixing of microscopic ensembles in smooth potentials. In such
experiments microscopic ensembles of regular orbits do not mix at all
in configuration space so all the observed mixing can be attributed to
chaos. However tests carried out  with ensembles selected by the above 
scheme in the  $N$-body isolated spherical NFW halo showed quite significant
spreading in configuration-space.  The evolution of the ensembles in
configuration space coordinates followed the following pattern:
ensembles spread rapidly in configuration space on a short timescale
(due to the Miller instability, see~\ref{sec:micro}), and then continued to spread more slowly
at a rate that scaled with local orbital period.  At a given radius
the mixing rate was found to be sensitive to the particular ensemble
in question - the nature of the parent orbit (whether it is box-like
or tube-like), the location of the orbit relative to the orientation
of the merger etc. Since their mixing in
configuration space is dominated by the Miller instability and phase
mixing, our ensembles are too large to be
usefully compared with the microscopic ensembles used to measure
chaotic mixing in the works of previous authors.  However, we found that ensembles selected by the above scheme in the
isolated halo showed negligible mixing in the coordinates $ E , {\mathbf J}$ (in agreement with
the previous findings of  \citeauthor{hut_heggie_01} \citeyear{hut_heggie_01}).
Consequently we focus our attention on mixing in these two
variables. The main objective of carrying out these experiments is to
observe mixing in $E,J$ that results from the merger.

We define the quantities (${\mathcal E}, {\mathcal J}$) for each particle in the ensemble as follows:
\begin{eqnarray} 
{\mathcal E} & = &   {\frac{E  - E(P^*)}{E(P^*)}}\\ 
{\mathcal J} & =  &  {\frac{J - J(P^*)}{J(P^*)}}
\end{eqnarray}
where $E(P^*), J(P^*)$ are the energy, magnitude of the total angular
momentum of the particle $P^*$ respectively. The quantities $\mathcal E, \mathcal
J,$ are computed for each of the 1000 particles in the ensemble.

The results of these experiments are described in \S~\ref{sec:mixing}.

\section{Results}

\subsection{Separation of Nearby Particles in Phase-Space \label{sec:separation}}

%
\begin{figure*}
\epsscale{0.9}
\plotone{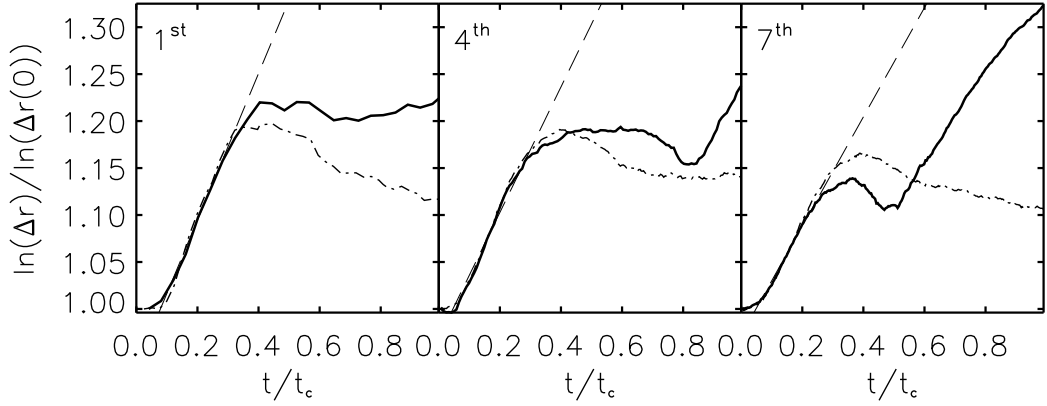}
\caption{Divergence in the quantities $\ln(\Delta r)/\ln(\Delta r
(t=0))$ as a function of time (in units of local crossing time) for
particles in the 1st, 4th and 7th radial shells.  The solid lines are
for the merger of two NFW halos (run Bp1). The dot-dashed lines are
for the control simulation of an isolated NFW halo. The dashed
straight lines are best-fits to both the solid lines and the
dot-dashed lines in the region $t/t_c = [0.05-0.35]$.}
\label{fig:miller}
\end{figure*}

We begin by demonstrating the ability of $\ln(\Delta r)$ to identify
chaotic behavior.  This quantity (and all the others that we use) are
compared to equivalent quantities measured in an isolated equilibrium
NFW halo. The three panels of Figure~\ref{fig:miller} show the initial
separation in spatial coordinate $\ln(\Delta r)$ (relative to its
value at $t=0$ Gyr) in three different radial shells (1st, 4th and
7th). The horizontal axis gives time in units of the median crossing
time $t_c$ in each shell of the corresponding isolated NFW halo.  The
local cross time are $t_c = 0.53, 2.59, 4.95$ Gyr for shells 1, 4, 7
respectively. We see that in both the merger (solid curve) and the
isolated halo (dot-dash curve) the initial separation is nearly linear
in the quantity $[\ln(\Delta r)/\ln(\Delta r(0))]$, indicating an
exponential change in radial separation.  The linear increase in
$\ln(\Delta r)$ on this timescale constitutes a clear manifestation of
the Miller instability (see \S~\ref{sec:micro}). This phase of
exponential divergence is short lived ($<$ 35\% of the local crossing
time). The $e$-folding time of the Miller instability $t_{e} = 0.39, 1.06, 
2.73$~Gyr for shells 1, 4, 7 respectively (roughly half the local crossing 
time in each shell).  In both the merger simulation and in the isolated halo,
the initial rapid change in $\ln(\Delta r)$ saturates at the same
value - a value determined by the length scale over which the smooth
potential dominates over the graininess \citep{kandrup_smith91}. At
smaller radii, the higher particle densities result in saturation
values of $\ln(\Delta r)$ that are higher  that  at larger radii.  Figure~\ref{fig:miller} is
clear evidence that micro-chaos is detectable despite our inability
(imposed by our use of a $N$-body distribution function) to choose
particles arbitrarily close to each other in phase space.

%
\begin{figure*}
\epsscale{0.9}
\plotone{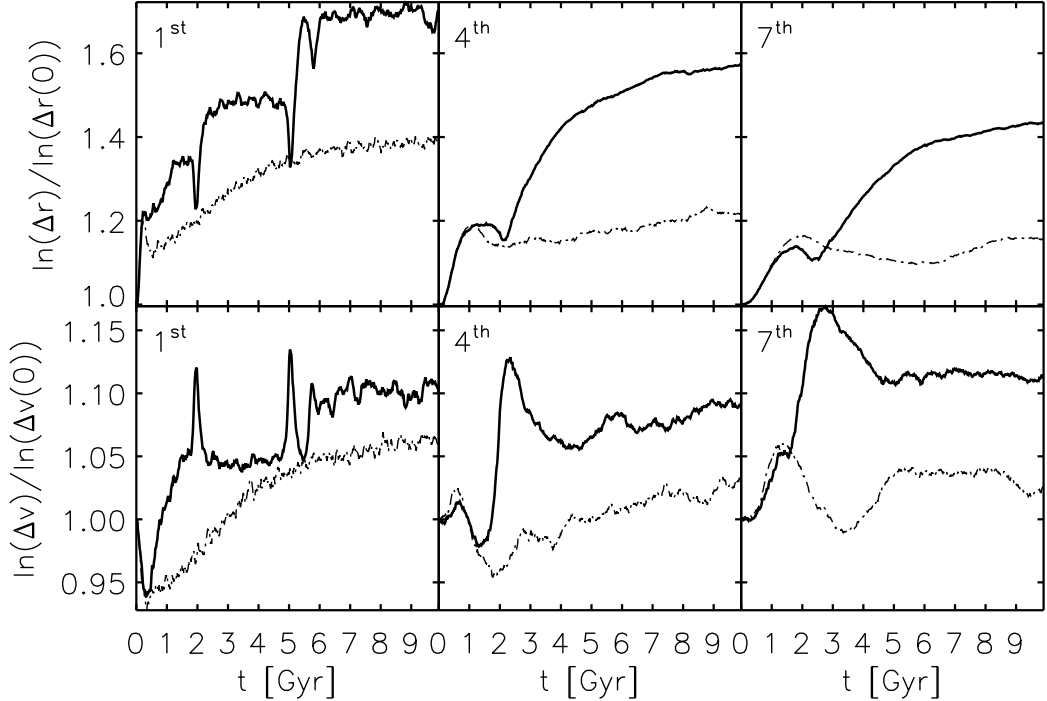}
\caption{The top three panels show divergence in the quantities
$\ln(\Delta r)/\ln(\Delta r (t=0))$ as a function of time for three
different shells in the halo (from left to right: the first, fourth,
and seventh shells respectively). The lower three panels show the
quantity $\ln(\Delta |v|)/\ln(\Delta |v (t=0)|)$, for the same three
shells. The solid lines are for the merger of two NFW halos (run
Bp1). The dot-dashed lines are for the control simulation of an
isolated NFW halo.}
\label{fig:rvsep}
\end{figure*}

We now focus on the behavior of the four quantities ($\ln(\Delta r)$, $\ln(\Delta |v|)$, $\Delta E$ and $\Delta J$) over longer time-scales. As before all quantities are compared with the equivalent quantities measured in an isolated equilibrium NFW halo to control for exponential Miller instability.  Figure~\ref{fig:rvsep} shows the evolution of $\ln(\Delta r)$ and $\ln(\Delta v)$ as a function of
time in 3 different radial shells (1st, 4th and 7th from the
center). The quantities plotted are scaled using the initial values of
the quantities at $t=0$ Gyr; i.e., the top panels show $[\ln(\Delta
r)/\ln(\Delta r(0))]$ and the lower panels show $[\ln(\Delta
|v|)/\ln(\Delta |v(0)|)]$ (where $\Delta r(0), \Delta v(0)$ are the separations at $t=0$).  
The solid lines show the evolution of
the quantities in the low-resolution merger simulations (run Bp1: two
NFW halos with $2\times 10^5$ particles per halo), and the dot-dashed
curves show the evolution of these quantities in the isolated
equilibrium $N$-body NFW halo. In what follows we point out some
 characteristic features of this figure that are seen elsewhere in
 this section.

The lower panels of Figure~\ref{fig:rvsep} show the separation in
$\ln(\Delta |v|)$ in both the isolated halo and the merging
halos. There is a noticeable initial decrease in $\ln(\Delta |v|)$
(which occurs simultaneously with the linear increase in $[\ln(\Delta
r)/\ln(\Delta r(0))]$), which we attribute to the mutual gravitational
attraction between the pair of particles that causes a deceleration of
the particles while their radial separation increases. The initial
decrease in $\ln(\Delta |v|)$ is seen in both the merging halos and
the isolated halo, confirming that this is not due to the merger but
is a characteristic of the $N$-body simulation.

In the innermost (1st) shell, after the initial growth in $\ln(\Delta
r)$ and the initial decrease in $\ln(\Delta |v|)$, there are periods of
time during which these quantities are essentially constant, separated
by several sharp decreases in $\ln(\Delta r)$ and correspondingly
sharp peaks in $\ln(\Delta |v|)$ not seen in the isolated halo. The
dips in $\ln(\Delta r)$ and peaks in $\ln(\Delta |v|)$ are coincident
with each other and correspond to the pericenter passages of the MBPs
of the two halos seen in Figure~\ref{fig:mbpvir}. It is the first and
second pericenter passages that cause the most significant changes in
the particle separations ($\ln(\Delta r$), $\ln(\Delta |v|$)).  

It is quite remarkable how little systematic  change in separation of
either $r$ or $v$ is seen in between pericenter passages, suggesting
that very little change in the macroscopic separations of particles
occurring at any time during the merger except during the pericenter
passages.  This can be understood as follows: during pericenter
passages of the centers of masses of the two halos, there is
significant overlap in the two potentials leading to a sudden
deepening in the potential experienced by the particles leading to a
transient {\it compressive} tidal field that simultaneously decreases
the separations in particles (seen by the dips in $\ln(\Delta r)$) and
increases their kinetic energies (seen in the sharp increases in
$\ln(\Delta |v|)$). Dynamical friction between
the two halos is also strongest at pericenter and consequently the
majority of their orbital energy and angular momentum are lost
impulsively at these times. When the two halos separate enough that their
innermost shells no longer overlap they experience little or no
change in their potentials, so there is little further change in separation of trajectories. 
Thus, the global potential fluctuations that occur between pericenter passages do not cause any
further increase in particle separations.  Behavior similar to that in shell 1 is
seen in shells 2 and 3 but is not shown here.

The effect of the first pericenter passage is seen to occur
simultaneously at all radii.  In shells 4, 7 we observe a dip in
$\ln(\Delta r)$ and peak in $\ln(\Delta |v|)$ at the first pericenter
passage which is both broader and less intense than in shell 1.  At
larger radii the first pericenter passage is followed by a continuous
change in $\ln(\Delta r)$ and $\ln(\Delta |v|)$ but subsequent
pericenter passages are less clearly visible.  This indicates that
while there is no propagation delay in the impulsive tidal shock with
radius, the response is longer lived at larger radii. This is because, the centers of the two merging halos
do not separate beyond the radius of the 4th and 7th shells after the first pericenter. Consequently the change in the external potential is not as sudden as for the inner shells.

%
\begin{figure*}
\epsscale{0.95}
\plotone{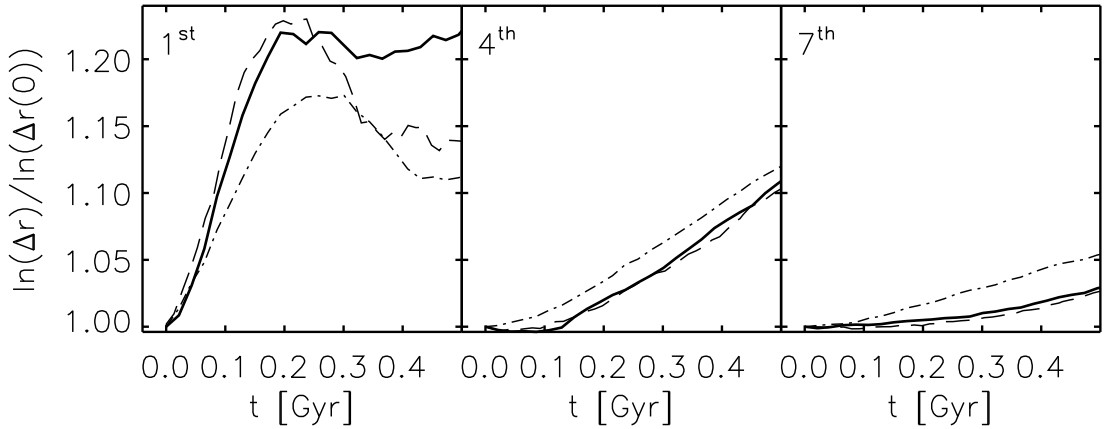}
\caption{The panels show divergence in the quantities $\ln(\Delta
r)/\ln(\Delta r (t=0))$ as a function of time for 1st, 4th, 7th
shells.  The solid curves are for the merger of two NFW halos (run
Bp1) with pairs of particles selected at $t=0.$~Gyr. The dashed curves
are for pairs of particles selected at $t=3.0$~Gyr in the isolated
halo and the dot-dashed curves are for pairs of particles selected at
$t=3.0$Gyr. The curves for the pairs starting at $t=3.0$~Gyr have been
translated in time to $t=0$~Gyr. The dot-dashed curves and the dashed
curves indicate that the initial exponential instability of nearby
orbits during the merger is not very different from that in the
isolated halo, independent of the time at which the particle pairs are
selected.}.
\label{fig:rvsep3gyr}
\end{figure*}

As discussed in \S~\ref{sec:definitions} it is extremely difficult to
make a reliable quantitative measurement of an exponential instability
on a short timescale. However, both the qualitative and quantitative behavior
of initially nearby particles seen in Figure 2 indicates that we are
in fact detecting the Miller instability. If the oscillating external
potential is also contributing to an exponential instability of orbits
one might expect that if pairs of nearby particles were picked during
the phase when the potential is changing very rapidly, the rate of
separation in some quantity (say $r$) would be greater than in either
the isolated halo or at the start of the merger.
Figure~\ref{fig:rvsep3gyr} compares the separations of particles
$\ln(\Delta r)$ when pairs were picked at $t=0$ Gyr 
with pairs picked at $t=0$ Gyr in the isolated halo  and
separation of pairs of particles picked at $t=3$~Gyr.
For the pairs of particles picked at $t=3$~Gyr in the merger, the time axis is $t-3$ so that initial slopes of the three
curves can be more readily compared. We chose a start time $t=3$~Gyr
to compare against since it corresponds to a time at which the virial ratio is changing
very rapidly and corresponds to the first apocenter separation of the
MBPs of the two halos (identical behavior was found when particles
were picked at $t=4$~Gyr). 

%
\begin{figure*}
\epsscale{0.95}
\plotone{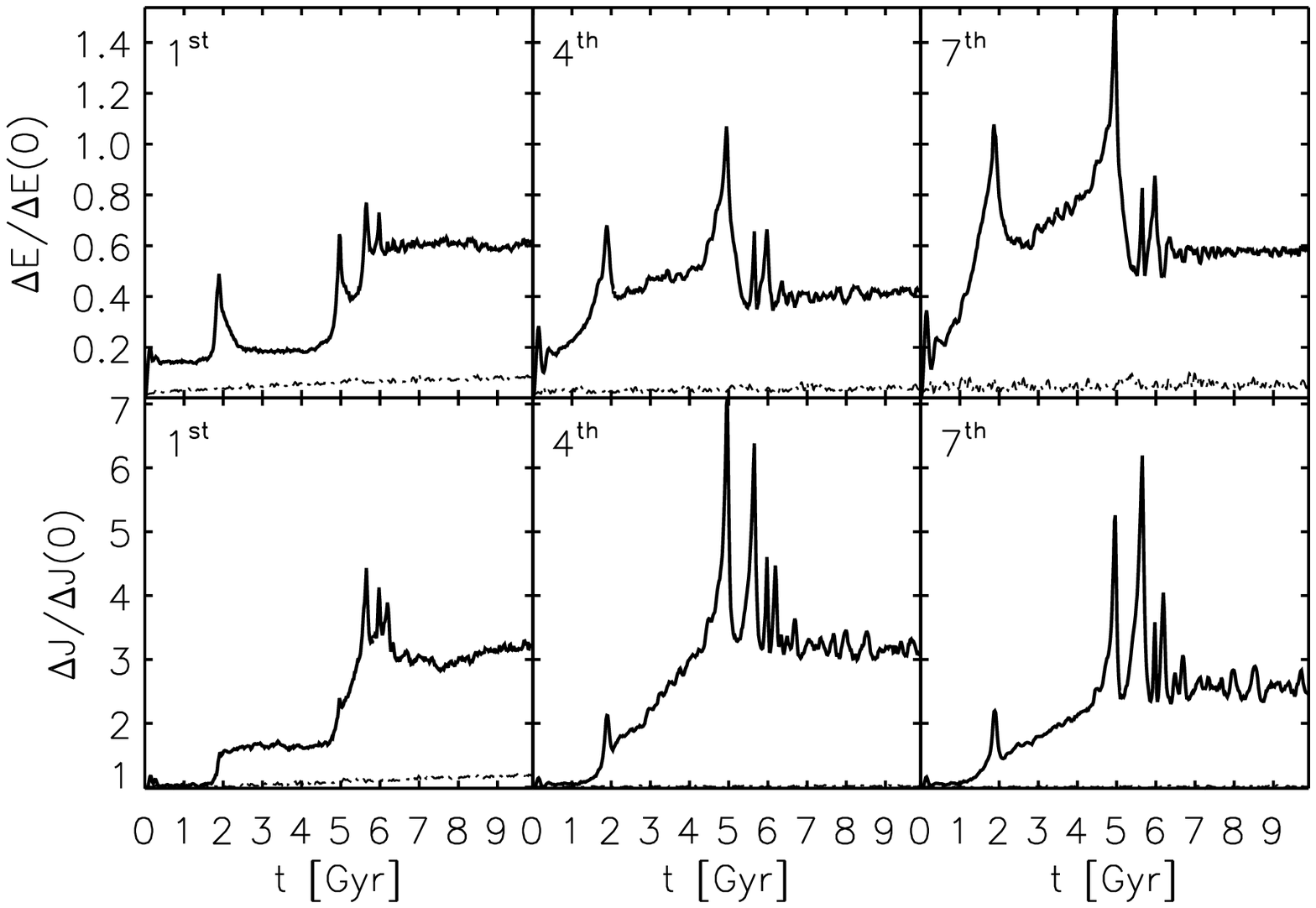}
\caption{The top three panels show divergence in the quantities
$\Delta E/\Delta E (t=0)$ as a function of time for three different
shells in the halo (from left to right: the first, fourth and seventh
shells respectively). The lower three panels show the quantity $\Delta
{\mathbf J}/\Delta {\mathbf J }(t=0)$, for the same three shells. The
solid lines are for the merger of two NFW halos (run Bp1). The
dot-dashed lines are for the isolated NFW halo.}
\label{fig:ejsep}
\end{figure*}

Since the potential is changing rapidly at
this time, we might expect to observe an enhanced rate
of separation of nearby pairs of orbits due to chaos induced by the
potential fluctuations  provided the $e$-folding time for this instability
is  comparable to or greater than that due to the Miller
instability. This figure shows, on the contrary, that the initial
exponential separation of particles in $r$ at $t=3$ Gyr  does not have a systematically greater slope than the rate of
separation of particles at $t=0$ Gyr. In shell 1 the
slope of the dot-dash curve is smaller than that of the solid curve
(which we attribute to the fact that heating from the first pericenter
passage caused nearby particles to be  further apart in
phase space); in shell 4 the slopes of all three curves are
essentially identical; only in shell 7 is the slope of the dot-dash
curve slightly steeper than that of the black curves indicating that
chaoticity may be playing a bigger role at large radii. 

While this is
clearly not conclusive evidence, it is suggestive of the hypothesis
that chaoticity due to the time-dependent potential, is  not
significant compared to the chaoticity due to the Miller instability. It
may play a small role in the evolution of the system at larger
radii.  However there is little evidence to suggest that
chaotic mixing from orbits in the time-dependent potential is {\em
driving} the relaxation.

To better understand the mechanism driving the relaxation we now look
at separation of particles in energy and angular momentum space. In
integrable time-independent potentials, energy and some quantity akin
to angular momentum are generally integrals of motion. It is therefore
of interest to quantify how these quantities change during a merger.

%
\begin{figure}
\epsscale{0.95}
\plotone{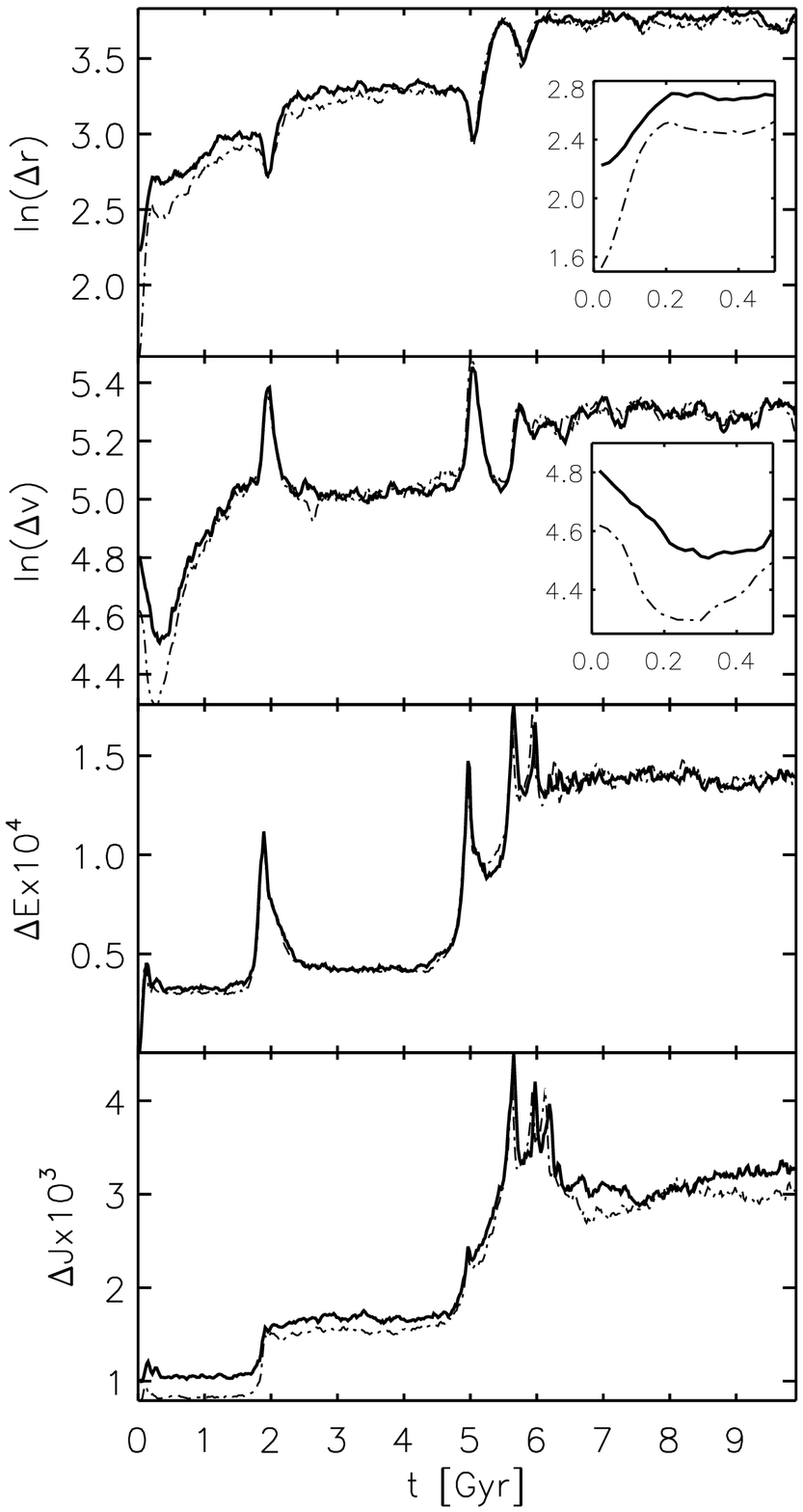}
\caption{The four panels (top to bottom) show divergence in the four
quantities $\ln(\Delta r)$, $\ln(\Delta |v|)$, $\Delta E$ and $\Delta
J$ plotted as a function of time for pairs of particles in the
innermost shell of one of the two merging NFW halos. The solid lines
are for the merger of two halos with $2\times 10^5$ particles in each
halo (run Bp1); dot-dashed lines are for the merger of two halos with
$2\times 10^6$ particles in each halo (run HRBp). The inserts in the
top two panels show differences in the initial separation due to the
Miller instability. The dot-dash line starts at lower values in both
$\ln(\Delta r)$, $\ln(\Delta |v|)$ which is a result of picking
particles closer in phase space in run HRBp.  For run HRBp $\ln(\Delta
r)$ has a shorter $e$-folding time, due to the higher particle number
and smaller softening length. Apart from the behavior of initial
separations in all four quantities, the low and high resolution
mergers are almost identical.}
\label{fig:rvejlohi}
\end{figure}

Figure~\ref{fig:ejsep} shows the separations in energies (top panels)
and angular momentum\footnote{The angular momentum ($J$) is about an
axis perpendicular to the plane containing the orbits of the
center-of-masses of the two merging halos.}  (bottom panels) for the
same pairs of particles as in Figure~\ref{fig:rvsep}. As before, the
dot-dashed lines are for pairs of particles in the isolated halo,
while the solid lines are for pairs of particles in one of the merging
halos. The quantities $\Delta E, \Delta J$ are scaled relative to
their initial values at $t=0$ Gyr. $\Delta E$ and $\Delta J$ (as
expected) are very close to zero in the isolated halo\footnote{We note that they are not precisely zero because, although
the total energy of the system is conserved to within numerical error,
the computation of $(\Delta E/\Delta E(t=0))$, $E$ for a given
particle is relative to the MBP - which can be a different particle at
each time-step.} .The sharp initial rise in $\Delta E$ and $\Delta J$ reflects the
response of the particles to the presence of second halo. This a
consequence of the fact that although each individual halo is in
equilibrium separately, when the two halos are set up on a parabolic
trajectory at $t=0$ they are no longer in equilibrium, since particles
now experience the external potential of the second halo.

After the initial increase in $\Delta E$ and $\Delta J$, there are
sharp but transient increases in the separation of these quantities
that also occurs primarily during pericenter passages. In fact, the
peaks in $\Delta E$ and $\Delta J$ are strongly correlated with the
pericenter passages in all radial shells and provide striking evidence
that it is these events that are primarily responsible for scattering
particles in $E$ and ${\mathbf J}$. We also note that the step-like
changes in these quantities and their correlation with pericenter
passages is seen at all radii unlike in the case of  $\Delta r$
and $\Delta v$ which changed more slowly and continuously in the outer
regions. In the inner most shells there is little or no separation in
phase-space quantities occurring between pericenter passages.  The
slight increase in $\Delta E$ and $\Delta J$ seen in shell 4 and shell
7 between $t=2$~Gyr (first pericenter passage) and $t= 5$~Gyr (second
pericenter passage) could be evidence for slower chaotic mixing.
 After the first pericenter passage the outer regions of the two
halos never completely separate again and the continued overlap of the
two potentials means that at larger radii the tidal compressive field
and dynamical friction is able to operate over a longer duration and
consequently more gradual changes occur as well. It is clear, however
that the impulsive changes in $\Delta E$ and $\Delta J$ during
multiple pericenter passages have the strongest effect on the mixing
in $E$ and ${\mathbf J}$.

%
\begin{figure*}
\epsscale{0.90}
\plotone{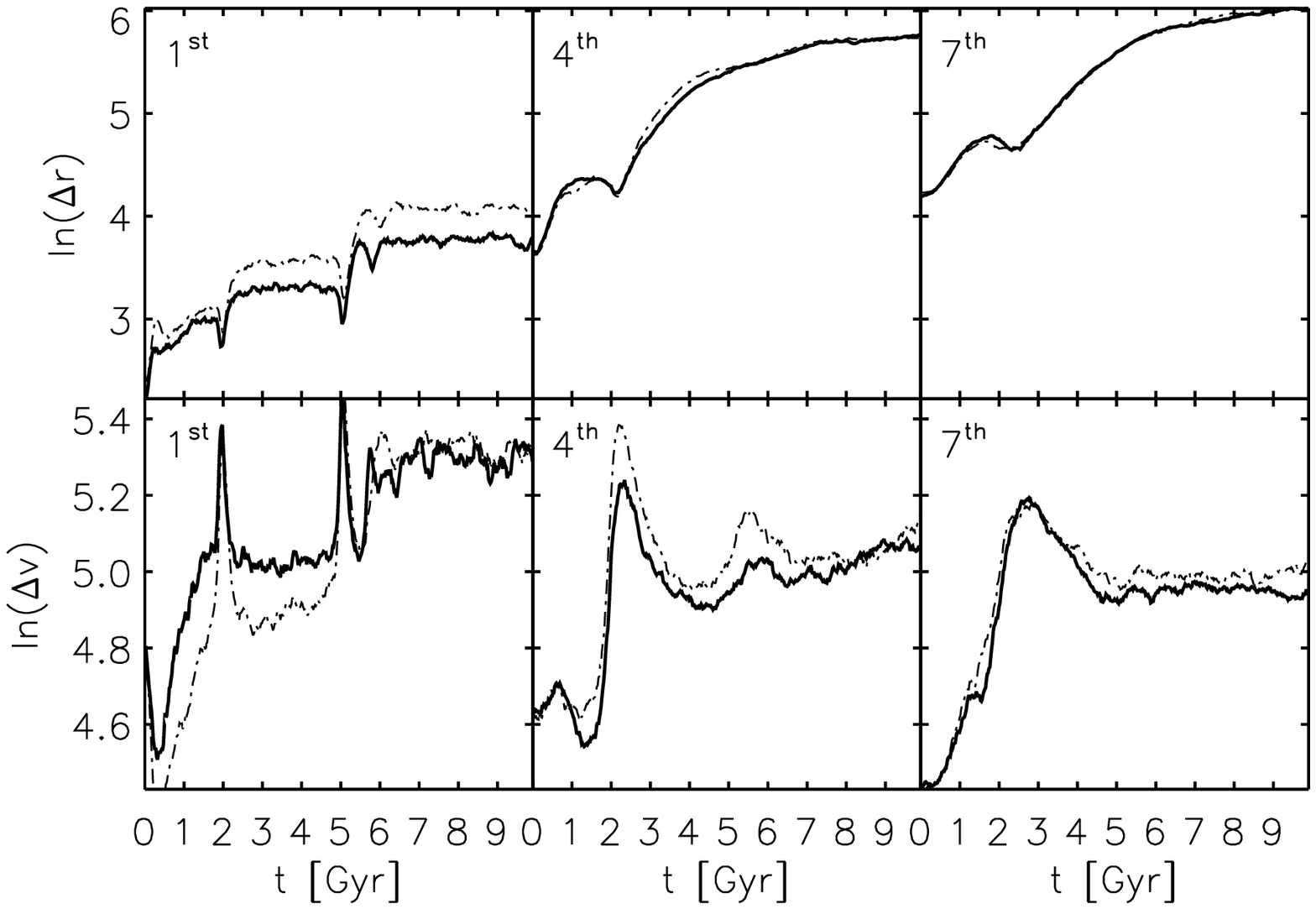}
\caption{Solid lines show separations in $\ln(\Delta r)$ and
$\ln(\Delta v)$ for the merger of two NFW halos and the dot-dash lines
show the separations of these quantities for the merger between an NFW
halo and a halo with an inner shallow cusp of $\gamma = 0.2$. The
behavior is similar in both cases except in the innermost shell where
the differences in the density profiles are significant.}
\label{fig:rvsepshallow}
\end{figure*}

Figure~\ref{fig:rvejlohi} compares separations in all four quantities
 $\ln(\Delta r)$, $\ln(\Delta |v|)$, $\Delta E$ and $\Delta J$ for two
 merger simulations with  different numerical resolutions. We recall that
 the run HRBp was performed with a factor of $10$ more particles and
 approximately a factor of $2$ higher force resolution than the run
 Bp1. We present the comparison for only the innermost shell since
 this is expected to be the most susceptible to numerical effects, if
 any.  There is clearly very little difference between the overall behavior of separations of particles in any of the 4 quantities plotted. The only noticeable difference is in the initial behavior of 
 $\ln(\Delta r)$, $\ln(\Delta |v|)$.
 
 The insert in the top most panel shows that the initial radial separation of
 particles in run HRBp is smaller at $t=0$ Gyr than it is for run
 Bp1. This is purely a consequence of the fact that by increasing the
 mass resolution the phase-space coordinates become more densely
 sampled and this allows us to select pairs of particles that are
 closer in phase-space (see inserts in top two panels). The insert in the second panel shows that the  
 initial velocity separation ($\ln(\Delta |v|)$) decreases to lower values in run  HRBp compared to run Bp1. This also  points to the increased gravitational deceleration due to the
 proximity of the nearest neighbor and the decrease in the softening
 parameter. 
 
 All curves are plotted without scaling them relative to
 their initial separations to illustrate that apart from differences
 in the initial separation (which we attribute to the Miller
 instability), the separations in all four quantities in the different
 resolution mergers saturate at the same values.  This is 
 confirmation that particle separations saturate at values that are
 determined by the global dynamics of the merger and not by the
 resolution of the simulations.

Figure~\ref{fig:rvsepshallow} shows separations in $\ln(\Delta r)$ and
$\ln(\Delta v)$ for particles in different radial shells in run Bp1
and in the merger of an NFW halo with a shallow cusp. The behavior is
remarkably similar in both cases except in the innermost shell. This
is expected, since the innermost shell is the only one in which the
density profiles (and consequently the phase-space density
distributions) differ significantly. It is well known from previous
studies \citep{boylan-kolchin_ma04,kazantzidis_etal06} that cusps in
shallow potentials are less robust than cusps in NFW halos. Particles
in the inner most shell of the shallow cusp experience a larger
increase in energy due to each pericenter passage than particles in
the NFW halo because they experience a greater relative increase in
the depth of the external potential arising from the overlap of the
two halos, and consequently the phase space volume accessible to them
is larger.  Beyond the scale radius the two halos have essentially
identical density profiles, and separations in $\ln(\Delta r)$ and
$\ln(\Delta v)$ are almost indistinguishable.

The results of our investigation of particle separations in $\ln(\Delta r)$, 
$\ln(\Delta |v|)$, $\Delta E$ and $\Delta J$  can be summarized as follows:
\begin{enumerate}
\item{The initial separation of pairs of nearby particles in $N$-body
systems such as those studied here is a result of the exponential
instability of the $N$-body problem, the so called ``Miller
instability.'' The qualitative and quantitative behavior of the
separation of particles: e.g. their dependence on local crossing time
(Fig.~\ref{fig:rvsep}), their dependence on softening and force
resolution (Fig.~\ref{fig:rvejlohi}) are completely consistent with
previous studies of this instability. It is important to point out
that the very fact that we are able to detect the exponential
divergence of orbits due to the Miller instability indicates that
despite the fact that pairs of nearby particles in the $N$-body
simulations are separated by macroscopic distances (i.e., are not
infinitesimally close by), our method for identifying an exponential
instability can be applied successfully to an $N$-body system.}
\item{Subsequent to the saturation of the separations due to the
Miller instability, the most significant increase in separations of
nearby particles (in $r$, $v$, $E$, or ${\mathbf J}$) occur during
pericenter passages of the MBP of the two halos and is a consequence
of the compressive tidal shocking (due to the overlap of the two
halos) and dynamical friction between the two halos that occurs during
the pericenter passages. In the innermost shells there is little or
no separation in phase-space quantities occurring between pericenter
passages, despite the large global fluctuations in potential occurring
during these times. At larger radii separation in $r$, $v$ continue
following pericenter passages.  At larger radii, there is  also a
small (but short lived) increase in separation in $E$ and $J$ between
the first and second pericenter passages.}
\item{If we consider the increase in the macroscopic separation of
particles in $(r, v)$ to be a measure of the mixing of particles in
these quantities, these plots would lead us to conclude that the
majority of mixing occurs due to the Miller instability and several
phases of mixing that occur during pericenter passages of the two
halos. In all radial shells of merger simulation we observe that the
macroscopic separation of initially nearby particles increases until
it saturates.  In the merging halos, saturation occurs at larger
values of $\ln(\Delta r)$, $\ln(\Delta |v|)$ respectively, than in the
isolated halo. This is a consequence of the fact that the accessible
phase-space volume has increased during the merger process.  We saw
from Figure~\ref{fig:ejsep} that $\Delta E$ and $\Delta J$ (which
represent increases in the available phase-space states accessible to
particles) increased in step-wise fashion primarily during pericenter
passages of the MPBs, as was predicted by \citet{spergelH92}.  By
$t=7$~Gyr the virial ratio $2T/|V| \approx 1$, although the potential
continues to experience long-lived but low-amplitude oscillations
until 13~Gyr. The low-level potential fluctuations cause little change
in the separation of either $r$ or $v$ beyond 7~Gyr.}
\end{enumerate}

In Lynden-Bell's (1967) model for ``violent relaxation'', particles
experience changes in their phase-space coordinates due to their
interaction with a time-varying background potential. It has been
argued by various authors that a time-dependent potential alone is
inadequate to cause relaxation, since nearby particles in phase space
will merely be relabeled in energy and will not separate (or
mix). The results from this section indicate that during the merger
of two halos, nearest neighbor particles in phase space do
mix in $E$ and ${\mathbf J}$ . Further, the primary causes of
mixing are the compressive tidal shocks and dynamical friction which
transfer energy and angular momentum from the orbital motion of the
two merging halos to the particles during the pericenter passages
\citep{spitzerChevalier73,gnedin99,valluri93}. The particles respond to sudden increments in
their orbital energy and angular momentum by jumping to new orbits,
resulting in mixing in phase space and evolution to a new
distribution. This latter phase of evolution could indeed be the
result of some weaker form of chaotic mixing, coupled with phase
mixing, both of which occur on a longer timescale.

\subsection {Mixing  of ensembles in phase-space  \label{sec:mixing}}

%
\begin{figure*}
\epsscale{0.9}
\plotone{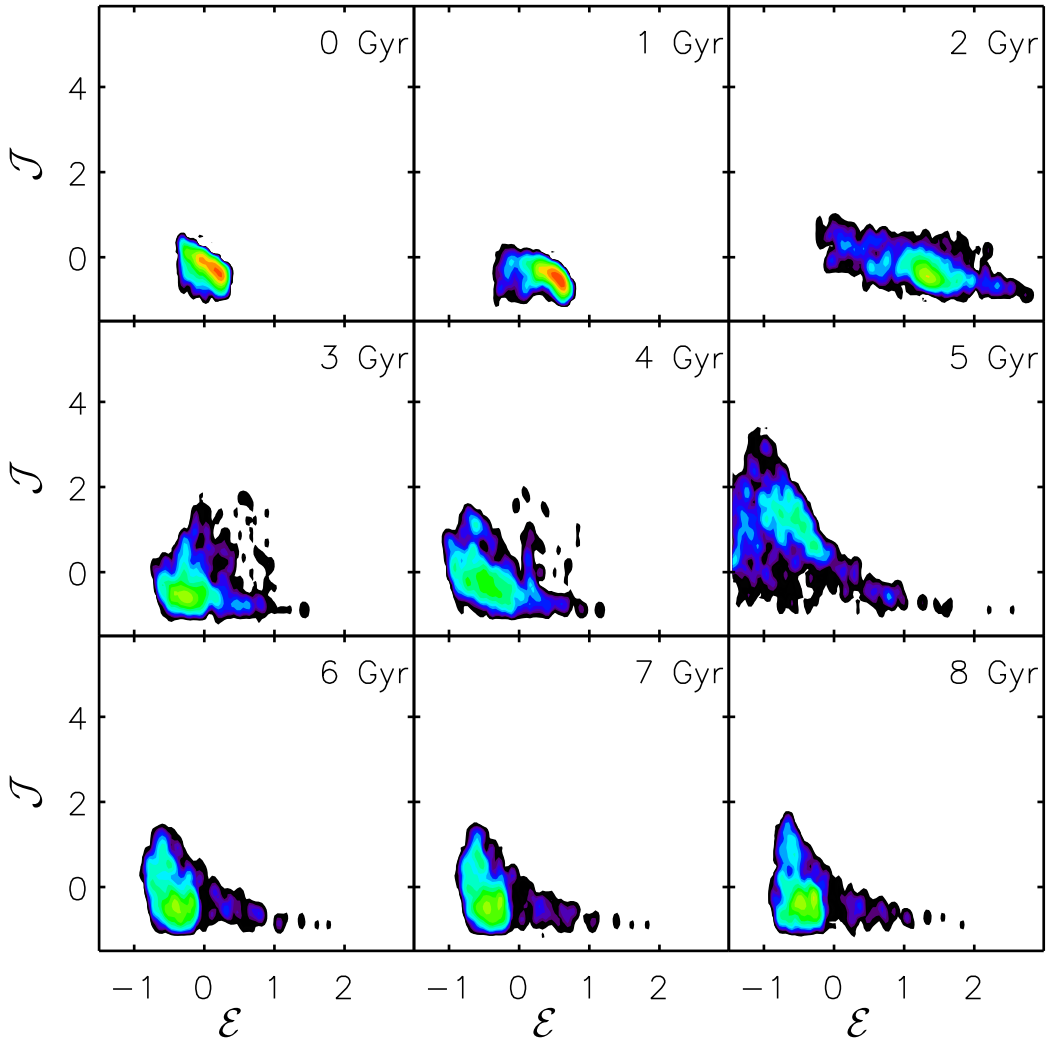}
\caption{The evolution of an ensemble of 1000 nearest neighbors in
phase-space in shell 4 at $t=0$ (run HRBp) plotted at 1 Gyr time
intervals in the merger.  The plot shows isodensity contours for the
particle distribution. The red contours correspond to the highest
density regions and black to the lowest density regions. The contours
are spaced at logarithmic density intervals relative to the maximum
density contour to enhance the visibility of low density
regions. Pericenter passages (at $t=2$ and $t=5$~Gyr) are seen to
cause a sudden spreading of the entire ensemble in ${\mathcal E}$ (a
result of the tidal shocking at these times). Two additional
pericenter passages occur between $t=5 and t=6$ (see Fig. 1). By
$t=6$~Gyr the ensemble has experienced 4 pericenter passages and has
settled to fill a triangular region in the ${\mathcal E}, {\mathcal
J}$ space.}
\label{fig:ejcontall}
\end{figure*}
In this section we present results of the mixing experiments with  an initially coherent
ensemble of  particles selected from the $N$-body simulation. As discussed in
\ref{sec:ensembles} we confine our analysis to mixing in the scaled  energy and angular-momentum  variables ${\mathcal E}, {\mathcal J}$.

Figure~\ref{fig:ejcontall} shows the evolution of an ensemble of 1000
self-gravitating particles in the 4th shell in run HRBp. The density
contours are obtained using a kernel smoothing algorithm. At $t=2$ and
$t=5$~Gyr which correspond to the first and second pericenter passage,
respectively, the ensemble undergoes a sudden spreading in energy
(${\mathcal E}$) which is a consequence of the deepening of the
potential during the overlap of the two halos.  By $t=6$~Gyr the
particle distribution fills a triangular region in ${\mathcal E},
{\mathcal J}$ space. A greater spreading in ${\mathcal E}$ at the
smallest values of angular momentum is a consequence of the fact that
these orbits are on the most radial orbits and consequently experience
greater changes in their potential energies. Particles with the
smallest (most negative) values of ${\mathcal E}$ experience the
greatest spread in angular momentum largely as a consequence of their
being ejected during the tidal shocks to larger radii.

We carried out similar mixing experiments for over 25 different
ensembles in each shell. The behavior of the ensemble in
Figure~\ref{fig:ejcontall} was found to be quite representative of all
the different ensembles. %

\begin{figure*}
\epsscale{0.9}
\plotone{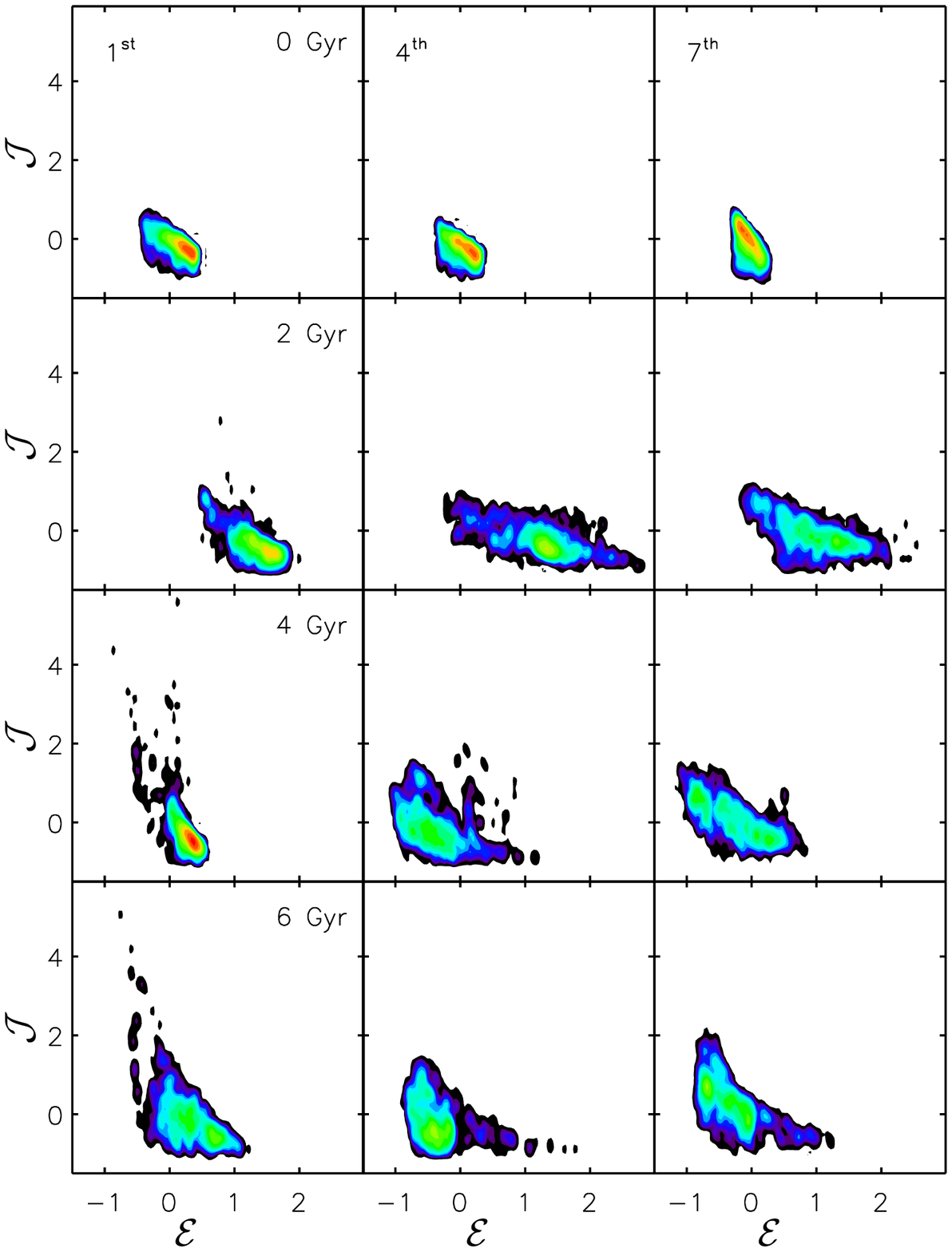}
\caption{Isodensity contours for ensembles of particles plotted as a
function of $\mathcal E, \mathcal J$ for shell 1(left hand column),
shell 4(middle column) and shell 7(right-hand column) plotted for
different times in their evolution (see text for details).}
\label{fig:ejcont}
\end{figure*}

Figure~\ref{fig:ejcont} shows contours of projected density of
particles as a function of the quantities ($\mathcal E, \mathcal J$)
in run HRBp, for ensembles in 3 different radial shells at 4
different times in the evolution. The evolution of the ensembles in
shells 1, 4 and 7 are very similar. At $t=2$~Gyr (first pericenter
passage) the ensemble spreads in $\mathcal E$. Similar increases in
spread and are seen during the 2nd pericenter passages (but are not
shown here).  After the first pericenter passage the majority of the
particles in the ensemble at $t= 4$~Gyr return to a more compact
distribution in ($\mathcal E, \mathcal J$) but one with a larger
spread in $\mathcal J$ at small $\mathcal E$. The bottom panels
correspond to $t=6$~Gyr, after the first three pericenter passages have
occurred. At this time all the ensembles fill a roughly triangular region in
$\mathcal E, \mathcal J$ space. This characteristic triangular final
distribution is seen in all shells and for essentially all of over a
hundred different ensembles we examined.  This final distribution is
not uniform in density, but the density contrast across the ensemble
is much smaller than in the initial distribution. We note that when
these ensembles were re-observed at $15$~Gyr  there was slightly greater uniformity of density in the
lower-density tails but otherwise they had changed very
little from their distributions at $8$~Gyr.

%
%

There are two important observations that can be made: (a) changes
 in $\mathcal E, \mathcal J$ occur
primarily during pericenter passages (Fig.~\ref{fig:ejcontall}) , and
(b) the final distribution in $\mathcal E,\mathcal J$ space of the
ensembles at all radii are similar, and the rate at which this final
distribution is reached is independent of radius (Fig.~\ref{fig:ejcont}).

The median crossing time in the innermost shell ($r\leq 25$~kpc) is
$t_c \sim 5\times 10^8$~yrs, while in the 7th shell is $t_c \sim 5$~Gyr. It
is obvious from Figure~\ref{fig:ejcont} that although the crossing
time in shell 7 is a factor of $10$ longer than in shell 1, the
diffusion of the ensembles in ($\mathcal E,\mathcal J$)-space appears
to occur primarily due to the pericenter passages of the MBPs and
there is only a slightly increased rate of spreading at larger radii.
Phase mixing and chaotic mixing in isolated potentials occur at rates
that scale with the local dynamical time so one would expect that
mixing effects that depend on the local dynamical time (such as
chaotic mixing) would occur faster at small radii than at large
radii. The fact that the evolution of the ensembles at different radii
occur episodically following pericenter passages, rather than
continuously at a rate that scales with local orbital time points to
the importance of the impulsive tidal events as the processes that
drives the transition to an equilibrium distribution in energy and
angular momentum.

Indeed, the absence of a strong dependence on local crossing time in
Figure~\ref{fig:ejcont} provides the strongest evidence so far that
 mixing in energy and angular momentum is driven by
compressive shocking and dynamical friction that occur at pericenter
passages. While chaotic mixing, as defined for static smooth
potentials, might be occurring during the merger of two halos, it
plays a minor role in driving the remnant to equilibrium. 

\section{Incomplete Relaxation and Redistribution of Particles in Mergers\label{sec:redistribution}}

Two interesting aspects of merger simulations of DM halos that have
been pointed out recently are (a) that cusps are remarkably robust and
(b) that about 40\% of the total mass of the remnant lies beyond
fiducial virial radius \citep{kazantzidis_etal06}.  Two questions that
arise from these findings are: 1) how do the final radial distributions, energies and angular momenta  of particles in the remnant depend on their original location in the merging halos and, 2) From where do the particles that end up outside the virial radius in the merger remnant originate? 
Addressing the first question is important since the merger causes the redistribution of particles 
to occur in a way that the density profiles and phase-space distribution functions preserve homology. This  may be interpreted as an indication that particles at all radii are
being heated uniformly at all radii. However, it is also reasonable to assume that less bound
particles in the outer part of the halo are preferentially heated.

It is also of interest to understand how the relative change in energy
or angular momentum of particles depends on their initial location in
the halo. In this section we examine the redistribution of particles
in radius, energy and angular momentum and the dependence of their
final locations on their initial location in the merging halos. We
note in passing that while the isolated halos and the initial merging
halos are spherical, justifying the use of spherical radial shells,
the merger remnant is significantly prolate-triaxial with
minor-to-major principal axis ratio $c/a$ varying from $0.5$ at the
center to $0.7$ at the virial radius. Although the merger remnants
exhibit significant departures from spherical symmetry, in what
follows we ignore the triaxiality of the final distribution when
binning particles in radius.

%
\begin{figure}
\epsscale{0.85}
\plotone{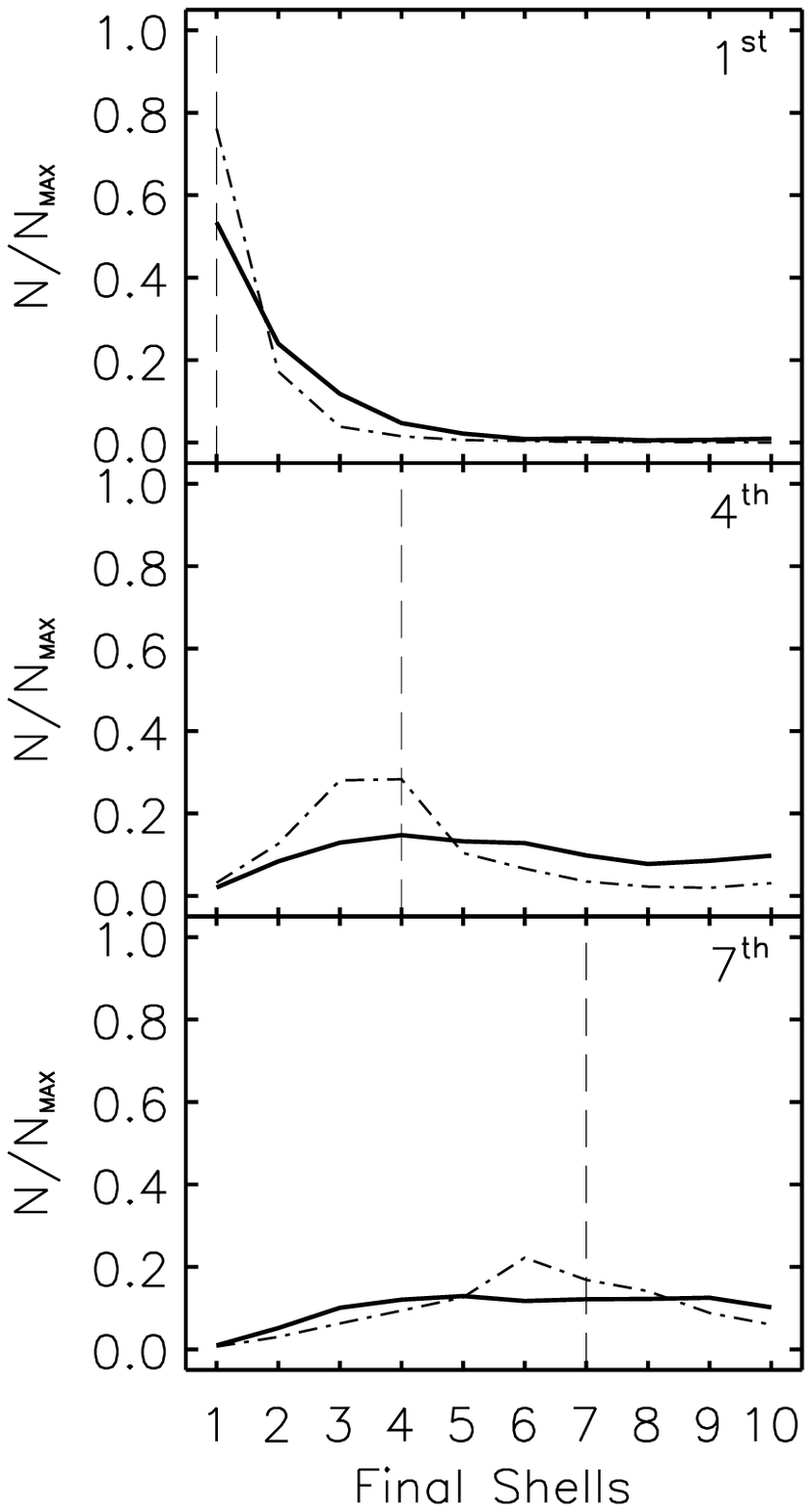}
\caption{The fraction of particles ($N/N_{\rm max}$) that lie in each
 of the radial shells at the final time step. The vertical dashed line
 and the caption in each panel indicate the shell in which particles
 were at $t=0$ Gyr. Particles from one of the two NFW halos (in run
 Bp1) is shown as a solid line while particles in the same shells of
 the isolated NFW halo are shown by dot-dashed line. }
\label{fig:rdistrib}
\end{figure}

Each panel in Figure~\ref{fig:rdistrib} shows how particles that were
originally in a given shell (indicated by the caption) are
redistributed in the final radial shells at $t=9$~Gyr when the merger
would conventionally be deemed complete. The results for isolated halo
are an important baseline for comparison.

In the top panel (shell 1), the solid lines (merger) and dot-dash
lines (isolated halo) are remarkably similar but indicate that while
the cusp is quite robust, in fact as much as 20\% of the particles in
the isolated halo and over 40\% of the particles in the merging halos
end up outside the cusp at the end of the simulation.  The primary reason 
is that 80\% of the particles in the
isolated halo have apocenters within the cusp.
 The merger  only causes a small additional fraction (20\%) of these bound 
 particles  to be spread to shells immediately
outside the cusp (shells 2-4).  In comparison
with the outer shells, this level of redistribution for shell 1 is
somewhat modest.

The middle panel (shell 4) shows that, for the isolated NFW halo
(dot-dashed curve), $\sim 50$\% of the particles  are in shells
3 and 4 at the end of the simulation, but the remainder are almost
uniformly redistributed within shells 2, and 5-10. In
shell 7  of the isolated halo $\sim 45$\% of the particles stay in shells 6 and 7  while the remainder are redistributed almost uniformly to all
radii from shell 4 outward. Thus, there is significant
radial redistribution of particles even in the non-evolving isolated
NFW halo purely because the initial distribution function has an
isotropic velocity distribution with roughly equal fractions of radial
and tangential orbits. The peaks in shells (3,4) and (6,7) are due to
the more tangentially biased orbits.  A comparison of the dot-dash curves
with the solid curves shows that the main effect of the merger is to
flatten out the peaks due to tangential orbits thus making the final
distributions more radially anisotropic.  At all radii the merger is
responsible for redistributing about $\sim 15-20 \%$ more particles from a given bin to
other radial bins.

A similar study of the merger of a NFW halo with a halo having a
shallow cusp showed that except in the innermost shell where the
profiles of the two halos differ, the particles in the two halos with
cusps of different slope are redistributed identically. A larger
fraction ($\sim 45$\%) of particles in the shallow cusp were redistributed by heating
to radial shells (2-4) directly outside the cusp (in comparison with 20\% in NFW-NFW halo mergers). The differences in
the distribution of particles at the end of the merger confirm that
NFW cusps are robust and retain a much more significant fraction of
their particles than do shallow cusps, confirming what has been
previously found by other authors
\citep{boylan-kolchin_ma04,kazantzidis_etal06}.

%
\begin{figure}
\epsscale{0.85}
\plotone{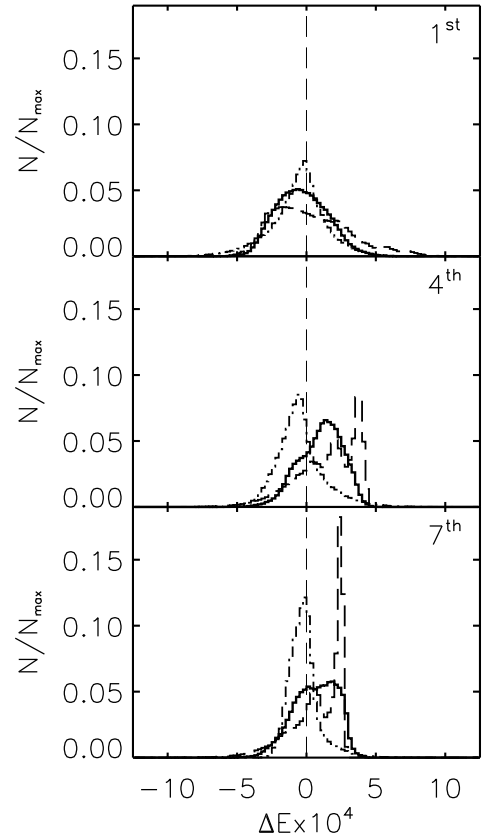}
\caption{Histograms of the change in energy ($\Delta E = E_{\rm final}
-E_{\rm initial}$) for all particles in shells 1 (top), 4 (middle),
and 7 (bottom).  The solid line shows the change in total energy, the
dot-dashed shows the change in kinetic energy, and the dashed line
shows the change in potential energy.}
\label{fig:ehist}
\end{figure}

Figure~\ref{fig:ehist} shows distributions of the changes in the
kinetic, potential, and total energies of all particles in different
shells at the end of the merger ($\Delta E = E_{\rm final} -E_{\rm
initial}$).  The y-axis gives the fractional change in number of particles (relative to the total number of particles in a given radial shell). Particles in the innermost shell experience a net
decrease in total energy that is a result of the overall decrease in
the potential energy of the particles pointing to an energy
segregation phenomenon during mergers \citep{funato_etal92}.  This is
expected, since there is an overall increase of $\sim 60$\% in the
mass of the cusp compared to the mass in the initial cusp
\citep{kazantzidis_etal06}. At all other radii, there is a net
increase in total energy that is most significant at larger
radii. This is a result of significant fractions of particles gaining
potential energy and being redistributed to larger radii as seen from
the multiple peaks in the potential energy distributions arising from
particles heated during each pericenter passage. It is striking to
note that in all shells the kinetic energy distributions remain peaked
about $\Delta E =0$ and are more peaked than Gaussian, indicating that
the majority of particles experience only a small change in their
kinetic energies despite experiencing large changes in their potential
energies.  This is additional confirmation that the redistribution in
energy that we saw in the mixing experiments (\S\ref{sec:mixing}) is
primarily due to changes in potential energies of particles due to
interactions with the background potential during pericenter passages.
%


\begin{figure}
\epsscale{0.85}
\plotone{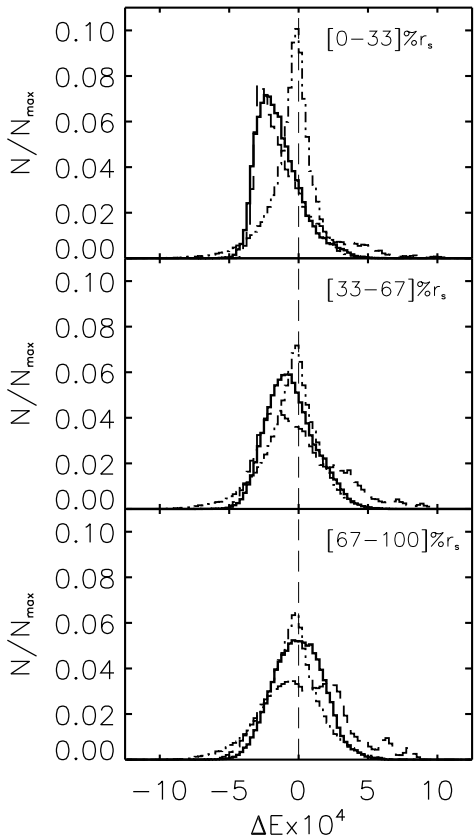}
\caption{Histograms of the change in energy ($\Delta E = E_{\rm final}
-E_{\rm initial}$) for all particles in three equal radius bins within
the scale radius ($r_s=21$ kpc). The solid line represents total
energy, the dashed-dot represents kinetic energy and the dashed
represents potential energy.}
\label{fig:ehistsc}
\end{figure}

The innermost shell includes all particles within $\sim 25$~kpc, which
is slightly larger than the initial scale radius ($r_s = 21$~kpc) of
the merging NFW halos. Figure~\ref{fig:ehistsc} is similar to
Figure~\ref{fig:ehist} but shows how particles in three equal radial
shells within $r_s$ are redistributed in energy.  It is clear that the
particles in the inner 33\% of the scale radius experience the
greatest deepening in the potential.

%
\begin{figure}
\epsscale{0.85}
\plotone{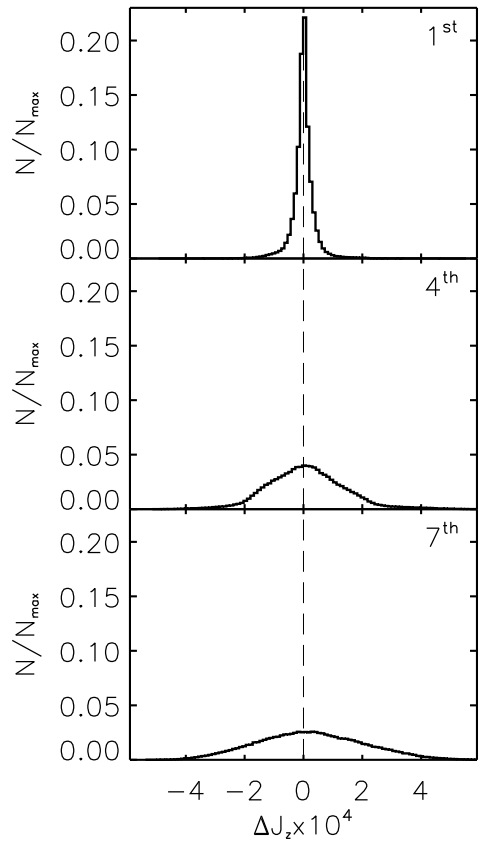}
\caption{Histograms of change in $J_z$ (component of ${\mathbf J}$
 perpendicular to the orbital plane of the merger) for all particles
 in the selected shells.}
\label{fig:jhist}
\end{figure}

Figure~\ref{fig:jhist} shows histograms of change in $J_z$ (the
component of ${\mathbf J}$ perpendicular to the orbital plane of the
merger) for all particles in radial shells 1, 4 and 7. The highly
peaked distributions in the innermost shell indicates that there is
virtually no net change in $J_z$ for the particles here.  In the outer
two shells, $\Delta J_z$ has an increasingly wider
distribution. Although it is not plotted here, we note that in the
outermost shells (shells 9, 10) the median of the distribution of
$\Delta J_z$ shifts to slightly positive values, indicating a small
gain in net angular momentum for the outermost particles. The observed
increase is a result of transfer of the orbital angular momentum of
the merging halos into the angular momentum of individual particles.
Our results indicate that the angular momentum of the merger is
absorbed primarily by particles at large radii.

\begin{figure}
\epsscale{0.9}
\plotone{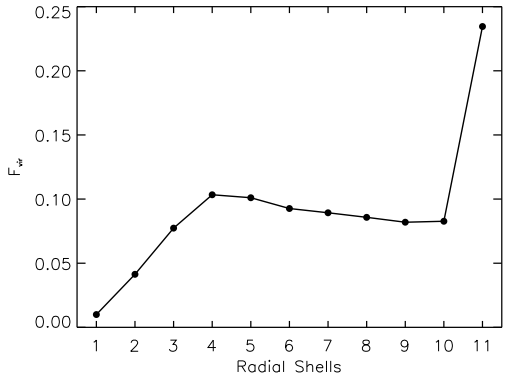}
\caption{Fraction of particles that originated in each of the radial
shells (1-11) in one of the original NFW halos involved in the merger
that end up outside the fiducial virial radius of the remnant. (All
particles beyond the virial radius are assigned to shell 11.)}
\label{fig:vir-out}
\end{figure}

As was noted previously, about 10\% of the mass of the initial DM
halos lies outside the fiducial virial radius, but it has been found
that nearly 40\% of the mass of the final remnant lies outside the
virial radius of the remnant halo \citep{kazantzidis_etal06}. We now
examine the distribution of particles in the merger remnant of run Bp1
to determine where these particles originate from.
Figure~\ref{fig:vir-out} shows the fraction ($F_{\rm vir}$) of the
total number of particles lying beyond the virial radius of the
remnant that originated from each of the $10$ shells of the original
halos.  All particles that lay outside the virial radius in the
original halos are assigned to shell 11 (which extends from the virial
radius to the outer edge of the simulation volume). The highest
fraction (about 25\%) of the particles outside the virial radius of
the remnant were already outside the virial radius of the initial
systems. The innermost shell is the most robust with fewer than ($<
1$\%) being ejected beyond $R_{\rm vir}$, Interestingly, all shells
from the half mass radius onward (shell 3 and beyond) contribute
roughly equally ($\sim 8-11$\%) to the particles that lie outside
$R_{\rm vir}$ in the final remnant. This figure and
Figure~\ref{fig:rdistrib} together show that the radial redistribution
of particles occurs essentially independently of the original radius
of the particle, provided that the particles lie outside the central
$3$ shells.

These results further highlight the fact that mass, in its standard
virial definition, is not additive in mergers. Models in which the
mergers double the mass within the virial radius of the remnant greatly
overestimate both the mass and density of the merger product.

\section{Summary and Discussion}

The principal goal of this study was to investigate the processes that
drive the evolution of self-gravitating systems to equilibrium and
cause mixing in phase-space. The goal was also to determine whether
the possible presence of large fractions of chaotic orbits arising
from the time-dependent potential was responsible for driving the
system to equilibrium.  We summarize the main results below.\\

%
%
%

\begin{enumerate}

\item{All orbits in the $N$-body simulation exhibit the
well-documented ``Miller instability'' of the $N$-body problem.  This
instability results in an initial nearly exponential separation of
nearby orbits in configuration space and energy, and a decrease in
velocity separation. The $e$-folding time ($t_e$) for the separation
increases with increasing softening length and decreasing particle
number. The initial exponential separation saturates on a short
timescale ($\sim 0.4 t_c$). The $e$-folding time and the number of
$e$-folds after which the instability saturates are identical in the
merging and isolated halos indicating that this is purely a
consequence of the $N$-body nature of the problem and not a result of
the merger. To this end, we use control simulations of isolated halos
as a baseline for comparison with merging halos to identify the mixing
processes occurring in mergers.}

\item{After the initial growth in $\ln(\Delta r)$ and decrease in
$\ln(\Delta |v|)$, there are periods of time for which these
quantities are almost constant, separated by several sharp
decreases in $\ln(\Delta r)$ and correspondingly sharp peaks in
$\ln(\Delta |v|)$ that are not seen in the isolated halo.  The
transient changes in $\ln(\Delta r)$ and $\ln(\Delta |v|)$ are
coincident with each other and correspond to the pericenter passages
of the centers of the merging halos.  Pericenter
passages also cause sharp transient step-like changes in the energy
and angular momentum separation of pairs of particles. Between
pericentric passages there is very little evidence for for mixing in the innermost shell,  but   at larger radii$\ln(\Delta r)$ and  $\ln(\Delta |v|)$ continue to grow slowly with time due to the overlap of the merging halos.}

\item{The changes in $\Delta E$ and $\Delta J$ are more strongly correlated  with pericenter passages at all radii pointing to the importance of compressive tidal shocks. At larger radii
there are is also some growth in separation between pericentric
passages, possibly pointing to the continued action of chaotic
mixing. If we assume that separations of nearby particles in phase-space
are linked to ``mixing'', this implies that the principal drivers of
strong mixing are the compressive tidal shock and increased dynamical
friction that occurs during the pericentric passages.}

\item{Mixing experiments with ensembles of 1000 nearest particles in
phase-space showed that mixing in the phase-space variables $\mathcal
E, \mathcal J$ also occurs primarily during pericenter
passages. Mixing leads  ensembles of nearby self-gravitating
particles to spread and fill a triangular region in $\mathcal E,
\mathcal J$ space. Mixing in these parameters occurs at the same
rate at all radii, largely independent of the characteristic crossing
time of the ensemble. This is further evidence that the mixing
process is driven by pericentric passages and is unrelated to mixing processes 
that scale with the local
crossing time. }


\item{We confirm the findings of several other authors that cusps of
DM halos are remarkably robust. The robustness of cusps is a
consequence of the fact that 80\% of the particles in the first shell
have apocenters that lie within the first shell. During the merger
only a small fraction (20\%) of particles with apocenters within the
original cusp are ejected to larger radii. The majority of the ejected
particles do not get beyond shell 4. A much larger fraction ($\sim$
45\%) of particles in the central regions of core-like profiles is
ejected to radial shells directly outside the cusp during a merger.}

\item{Particles in a given radial shell outside the cusp are
redistributed almost uniformly with radius primarily due to phase
mixing. This result holds for both the isolated spherical halo and
the merging halos. The merger helps to redistribute an additional $\sim
$20\% of orbits that have predominantly tangential velocity
distributions in the original halos. }

\item{Particles within the scale radius of the merging NFW halos
experience a net decrease in total energy following the merger. This
is a result of the overall decrease in the potential energy of the
particles, resulting from the ($\sim 60$\%) increase in the mass of
the cusp in the final remnant.}

\item{The majority of particles in the merging halos experience only a
small change in their kinetic energies and moderate 
changes in their potential energies. The change in total energy of
particles is driven predominantly by the changes in their
potential energies during pericenter passages. }

\item{At smaller radii the majority of particles experience only small
change in angular momentum evidenced by highly peaked distributions of
$\Delta J$. The angular momentum of the merger is absorbed primarily
by particles in the outermost radial shells. }

\item{The largest fraction of particles lying beyond the virial radius
of the final remnant (25\%) were located beyond the virial radius of
the original halos. The inner 2 shells are most robust to the ejection
of particles beyond the virial radius. All shells between shell 4 and the virial radius (shell 10) contribute
roughly equally to the particles ejected beyond $R_{\rm vir}$ of the
remnant.}
\end{enumerate}

While this paper focuses on simulations of the merger of two dark
matter halos with NFW potentials, most of the results are generic to
mergers of collisionless gravitating systems and therefore applicable
to dissipationless (``dry'') mergers of elliptical galaxies.  In
particular, the absence of large amounts of mixing in radius over and
above that expected from phase mixing in an isolated halo supports
previous work that indicates that radial gradients in stellar
properties (such as metallicity gradients) are can to survive intact
even in major mergers \citep{white80,barnes96,boylan-kolchin_ma04}.
These results are also expected to be generically applicable to
cosmological mergers. In fact soon after the submission of this paper
we became aware of the work of \citet{faltenbacher_etal06} - a study
of the relaxation processes that operate in hierarchical mergers of
dark matter halos in cosmological $N$-body simulations. These authors
independently arrived at the same conclusions that we do, namely that
the primary driver for mixing and relaxation in cosmological mergers
is tidal shocking.

Finally, we find strong evidence that the processes that drive the
mixing in phase space and evolution to a dynamical equilibrium do not
occur continuously due to the time-dependent potential of the
system. In fact strong mixing occurs primarily following episodic
injection of energy and angular momentum into the internal energies of
particles during pericenter passages. The injection of energy and
angular momentum causes nearest particles in phase-space to separate
in $E, {\mathbf J}$.

Previous studies of orbital evolution in time-dependent and
time-independent potentials have argued that the presence of large
fractions of chaotic orbits could be the principal driver of the
mixing and evolution to an equilibrium distribution.  While chaotic
mixing could well be occurring, we do not find evidence that such
chaotic mixing is driving the relaxation. This is probably a consequence of the
fact that the timescales required for chaotic mixing to be relevant
are much longer than the duration of the merger. 

Two  interesting questions remain regarding ``violent relaxation'' 1) does the mixing that occurs during collapse of an isolated cold  gravitating system give rise to similar mixing in phase-space? 2) what gives rise to the universal power-law phase-space density profiles seen in cosmological mergers? We defer addressing both questions to future work.

\acknowledgments

We would like to acknowledge the inspiration provided by the late
Henry Kandrup (who was IMV's thesis advisor at the time of his death)
and his significant contributions to this subject.  This work has been
significantly influenced by his legacy. We thank D. Merritt,
R.H. Miller, I.V. Sideris, C. Siopis, S. Tremaine, P. Vandervoort and
T. de Zeeuw for detailed comments on an earlier draft of this
paper. MV is supported by the Kavli Institute for Cosmological Physics
(KICP) at the University of Chicago.  IMV acknowledges the support of
National Science Foundation (NSF) grant AST-0307351 to the University
of Florida and the support of Northern Illinois University and KICP.
SK is supported by the Swiss National Science Foundation and by the
KICP. AVK is supported by NSF under grants AST-0507666 and
AST-0239759, and by NASA through grant NAG5-13274. CLB is supported by
the Department of Energy through grant DE-FG02-04ER41323 to NIU. This
work used the resources and support of KICP which is funded by NSF
through grant PHY-0114422.  The numerical simulations used in this
study were performed on the zBox1 supercomputer at The University of
Z\"urich. This research made use of the NASA Astrophysics Data System.

\bibliographystyle{abbrvnat}
\bibliography{ms}

\end{document}